\documentclass[12pt]{article}
\pdfoutput=1
\usepackage{graphicx}
\usepackage{color}
\usepackage{amssymb,amsmath,bm}
\usepackage{epsf}
\usepackage{epsfig}
\usepackage{afterpage}
\usepackage{longtable}
\usepackage{cite}
\usepackage{latexsym,mathrsfs}
\usepackage{graphics}
\usepackage{url}
\usepackage{paralist}
\usepackage{subcaption}
\usepackage{psfrag}

\newcommand{\LinkColor}[0]{black}

\usepackage{hyperref} % references
\hypersetup{
	colorlinks, 
	bookmarksopen, 
	bookmarksnumbered,
	pdftoolbar=false, % show Acrobat toolbar
	pdfmenubar=false, % show Acrobat menu?
	pdffitwindow=false, % window fit to page when opened
	pdfstartview={FitH},	 % fits the width of the page to the window
	citecolor=\LinkColor, %NavyBlue, 	%color of links to bibliography
	linkcolor=\LinkColor,	%color of internal links
	urlcolor=\LinkColor, %color of external links 
	pdfnewwindow=true, % links in new window
	linktoc=all
}

\usepackage{float}

\newcommand{\B}{{Z'}}
\usepackage{booktabs}
\definecolor{red}{cmyk}{0,1,1,0.4}
\definecolor{darkgreen}{rgb}{0.0,0.6,0.0}
\definecolor{cDarkGrey}{RGB}{91,91,91}
\definecolor{cGrey}{RGB}{245,243,238}
\definecolor{cBlue}{RGB}{0,110,191}
\definecolor{cLightBlue}{RGB}{214,237,252}
\definecolor{cRed}{RGB}{196,0,100}
\definecolor{cLightRed}{RGB}{254,222,237}
\definecolor{cGreen}{RGB}{0,166,80}
\definecolor{cLightGreen}{RGB}{254,222,237}
\definecolor{cOrange}{RGB}{221,74,44}
\definecolor{cLightOrange}{RGB}{255,215,210}
\definecolor{cPurple}{RGB}{93,35,125}
\definecolor{cLightPurple}{RGB}{241,230,252}
\definecolor{cYellow}{RGB}{252,191,10}
\definecolor{cISSRBlue}{RGB}{0,111,174}
\definecolor{cISSRGrey}{RGB}{167,169,172}

\newcommand{\ord}{\mathcal{O}}

\newcommand{\IM}{\text{Im}}
\newcommand{\RE}{\rm{Re}}
\newcommand{\imlt}{\IM\lambda_t}

\newcommand{\tev}{\, {\rm TeV}}
\newcommand{\gev}{\, {\rm GeV}}

\def\B{\mathcal{B}}
\newcommand{\vcb}{|V_{cb}|}
\newcommand{\vtd}{|V_{td}|}
\newcommand{\vub}{|V_{ub}|}
\newcommand{\vts}{|V_{ts}|}
\newcommand{\vus}{|V_{us}|}

\def\ksm{K_S\to\mu^+\mu^-}
\def\klm{K_L\to\mu^+\mu^-}

\newcommand{\bei}{B_8^{(3/2)}}
\newcommand{\kepe}{\kappa_{\varepsilon^\prime}}

\def\klpll{K_L\to\pi^0\ell^+\ell^-}
\def\epe{\varepsilon'/\varepsilon}
\newcommand{\beq}{\begin{equation}}
\newcommand{\eeq}{\end{equation}}
\newcommand{\be}{\begin{equation}}
\newcommand{\ee}{\end{equation}}
\newcommand{\bi}{\begin{itemize}}
\newcommand{\ei}{\end{itemize}}
\newcommand{\ba}{\begin{array}}
\newcommand{\ea}{\end{array}}
\newcommand{\beqa}{\begin{eqnarray}}
\newcommand{\eeqa}{\end{eqnarray}}
\newcommand{\bea}{\begin{eqnarray}}
\newcommand{\eea}{\end{eqnarray}}
\newcommand{\beqn}{\begin{eqnarray}}
\newcommand{\eeqn}{\end{eqnarray}}

\newcommand{\BR}{{\cal B}}

\newcommand{\lsim}{\stackrel{<}{_\sim}}
\newcommand{\gsim}{\stackrel{>}{_\sim}}

\definecolor{red}{cmyk}{0,1,1,0.4}

\newcommand{\kpn}{{K^+ \to \pi^+\nu\bar\nu}}
\newcommand{\klpn}{{K_L \to \pi^0\nu\bar\nu}}

\usepackage{fancyhdr}
\renewcommand{\sectionmark}[1]{
  \markboth{\thesection\ ~~#1}{\thesubsection\ #1}}   % section title
\advance \headheight by 3.0truept       % for 12pt mandatory...
\usepackage[left]{lineno}
\begin{document}

\begin{flushright}
    {AJB-26-3}\\
    P3H-26-058\\
    TTP26-030
\end{flushright}

\medskip

\begin{center}
{\Large\bf
\boldmath{Prospects for $\klpn$, $K_S\to \mu^+\mu^-$, $K_L\to\pi^0 \ell^+\ell^-$ and $\epe$  after the new $\kpn$ result from NA62}}
\\[0.8 cm]
{\bf Monika Blanke$^{a,b}$,  
Andrzej~J.~Buras$^{c,d}$,
Cristina Lazzeroni$^{e}$ and  
Joel C. Swallow$^{f}$
\\[0.5 cm]}
{\small 
$^a$Institut f\"ur Astroteilchenphysik, Karlsruhe Institute of Technology, Hermann-von-Helmholtz-Platz~1, D-76344 Eggenstein-Leopoldshafen, Germany\\
$^b$Institut f\"ur Theoretische Teilchenphysik, Karlsruhe Institute of Technology, Engesserstra{\ss}e 7, D-76128 Karlsruhe, Germany\\
$^c$TUM Institute for Advanced Study, Lichtenbergstr.~2a, D-85748 Garching, Germany\\
$^d$Physik Department, TUM School of Natural Sciences, TU M\"unchen,\\ James-Franck-Stra{\ss}e, D-85748 Garching, Germany\\
$^e$University of Birmingham, Edgbaston, Birmingham B15 2TT, United Kingdom\\
$^f$CERN, Esplanade des Particules 1, 1211 Meyrin, Switzerland\vspace{3mm}\\ 
E-mail: {\tt monika.blanke@kit.edu}, {\tt aburas@ph.tum.de},\\ {{\tt cristina.lazzeroni@cern.ch}, {\tt joel.christopher.swallow@cern.ch}}}
\end{center}

\vskip0.41cm

\noindent
\begin{abstract}
The recently announced NA62 measurement of the $\kpn$ branching ratio, based on 2016--2024 data, is fully consistent with its very accurate Standard Model (SM) prediction. 
While it is not excluded that the final result based on 2016--2026 data will deviate from the SM prediction, the question arises whether a similar fate awaits the $\klpn$ decay.
This decay is currently being searched for by the KOTO experiment, with the present upper bound roughly two orders of magnitude above its very precise SM prediction. 
The proposed KOTO II experiment aims to provide the first discovery of the $\klpn$ decay and measure its branching ratio.
Building on the findings of several previous papers we demonstrate that in the presence of suitably chosen large new complex phases and of a small amount of new right-handed $\bar s d$ couplings, in addition to the left-handed ones, the $\klpn$ branching ratio can still be enhanced by one order of magnitude with respect to the SM prediction while keeping both $\kpn$ and $\varepsilon_K$ SM-like. 
Simultaneously the decays $K_S\to\mu^+\mu^-$,  studied by LHCb, and $K_L\to\pi^0\ell^+\ell^-$,  searched for by KOTO II, can be strongly enhanced and the anomaly in the ratio $\epe$, as claimed by Dual QCD, removed. We illustrate this with an example of a specific $Z^\prime$ scenario. 
\end{abstract}

\thispagestyle{empty}
\newpage
\tableofcontents
\newpage
\setcounter{page}{1}

%%%%%%%%%%%%%%%%%%%%%%%%%%%%%%%%%%%%%%%%%%%%%%%%%%%%%%%%%%%%%%%%%%%%%%
\section{Introduction}
Rare decays of kaons have, 
for decades, played a very important role in testing the Standard
Model (SM) and in the search for new physics (NP). Recent reviews of these decays can be found in \cite{Aebischer:2022vky,Aebischer:2025mwl}. Table 2 in \cite{Aebischer:2022vky} demonstrates in an impressive manner the interest in these decays in the context of the search for NP.

The main players in this game are the decays
\be\label{OBS}
\kpn,\quad \klpn,\quad K_L\to \mu^+\mu^-,\quad K_S\to\mu^+\mu^-,\quad  K_L\to \pi^o\ell^+\ell^- ~(\ell = e,\mu)\,
\ee
as well as the two CP-violating observables
\be \label{CP}
\varepsilon_K, \quad \epe.
\ee
The status of these six rare decays and of the two observables, summarised  recently in \cite{Buras:2026hbe}, is as follows:
\begin{itemize}
\item
The NA62 collaboration at CERN 
recently published a measurement of the $\kpn$
branching ratio based on 2016--2024 data with the result~\cite{NA62pnn2026}~\footnote{The impressive progress made by the NA62 collaboration on $\kpn$ and by KOTO on $\klpn$ becomes evident when reading the extensive review on the status and perspectives on both decays written in 2015 just after the start of the NA62 experiment~\cite{Buras:2015qea}.}
\be\label{EXP19}
\mathcal{B}(\kpn)_\text{exp}=(9.6^{+1.9}_{-1.8})\times 10^{-11}\,.
\ee
This is fully consistent with the SM estimates given below in (\ref{BV26}) and
(\ref{GSV}) but still leaving room for some NP contributions.
The NA62 experiment will complete collecting its full dataset in 2026, with the 2025--2026 data---amounting to approximately $1/3$ of the full 2016--2026 dataset---still to be analysed.
\item
The most recent
$90\%$ confidence level (C.L.) upper bound on $\klpn$ from KOTO~\cite{KOTO:2024zbl} reads
\be\label{EXP19L}
\mathcal{B}(\klpn)_\text{exp}\le 2.2\times 10^{-9}\,,
\ee
roughly two orders of magnitude above the SM predictions given in (\ref{BV26}) and (\ref{GSV}). 
The KOTO experiment will improve its sensitivity by an order of magnitude by 2030, 
and ``an indication of new physics with a significance of $90\%$ is possible if the observed branching ratio differs by $40\%$ from the SM value''~\cite{KOTO:2025uqg}. 
The proposed KOTO II experiment~\cite{KOTO:2025gvq} targets the first discovery of the $\klpn$ decay and a measurement of its branching ratio in the late 2030s.
\item 
The decay $K_L\to\mu^+\mu^-$ is subject to significant long-distance contributions. Nonetheless an experimental bound on the short-distance (SD) contribution, sensitive to NP, can be derived~\cite{Isidori:2003ts}:
\be
\mathcal{B}(K_L\to\mu^+\mu^-)_{\rm SD} \le 2.5 \times 10^{-9}\,.
\ee
\item
The present upper bound on $\ksm$ from the LHCb collaboration, using data from 2011--2012 and 2016--2018 (Run1 + Run2), reads~\cite{LHCb:2020ycd}
\begin{align}
  \label{ksmbound}
   \BR(\ksm)_{\rm LHCb} & < 2.1 \times 10^{-10} \; \mbox{at} \; 90\% \; \mbox{C.L.} \,.
   \end{align}
This result is about a factor of $50$ above
the SM prediction given in (\ref{ISDA}) \cite{Isidori:2003ts, DAmbrosio:2017klp}. LHCb expects to significantly improve the sensitivity using Run3 data, with strong prospects to make a measurement of the branching ratio by the end of LHC operations~\cite{AlvesJunior:2018ldo}. 
\item 
  The present experimental bounds on the decays $K_L\to \pi^o\ell^+\ell^-$ from  the KTeV experiment are  
\be
%\label{KLLL}
{\cal B}(K_L\to\pi^0e^+e^-)<28\times10^{-11}\quad\text{\cite{AlaviHarati:2003mr}}\,,\quad
{\cal B}(K_L\to\pi^0\mu^+\mu^-)<38\times10^{-11}\quad\text{\cite{AlaviHarati:2000hs}}\,,
\ee
which are still by one order of magnitude larger than the respective SM predictions.
The KOTO II proposal includes plans to measure the branching ratios of these decays~\cite{KOTO:2025gvq}. 
\item
The $K^0-\bar K^0$ CP-violating parameter $\varepsilon_{K}$ is fully consistent with the SM calculations, as  emphasised in \cite{Buras:2021nns,Buras:2022wpw}.\footnote{Note however the strong dependence of $\varepsilon_{K}$ on the CKM parameter $\vcb$, calling for a solution of the tensions in the determination of the latter parameter.}
\item
  The ratio $\epe$ quantifying the amount of direct CP violation in $K_L\to\pi\pi$ decays, as measured by the NA48 collaboration at CERN \cite{Batley:2002gn} and KTeV at Fermilab \cite{AlaviHarati:2002ye,Abouzaid:2010ny},
  is at least a factor of two above the SM prediction obtained using the Dual QCD approach \cite{Buras:2015xba}. 
    While the results from the RBC-UKQCD collaboration \cite{Abbott:2020hxn} using lattice QCD methods and from Chiral Perturbation Theory \cite{Cirigliano:2019ani,Gisbert:2020wkb} are consistent with the experimental value, large uncertainties do not allow definite conclusions to be drawn.    
    Recent reviews on this topic can be found in \cite{Buras:2020wyv,Aebischer:2022vky}.
  \end{itemize}
As evident from many papers listed in Table~2 of \cite{Aebischer:2022vky} and
in Table~8 of \cite{Buras:2026vbp}, the observables listed in (\ref{OBS}) and \eqref{CP}
can, in certain NP scenarios, be strongly correlated with each other. 
A recent SMEFT analysis in the context of a particular $Z^\prime$ model presented in
\cite{Aebischer:2023mbz} demonstrated this in an impressive manner:
assuming the flavour-changing
$\bar s d Z^\prime$ coupling to be purely left-handed (LH) and imaginary, a possible enhancement of the
$\kpn$ branching ratio implies enhancements of the rare decays $\klpn$, $K_S\to\mu^+\mu^-$, $K_L\to\pi^{0}\ell^+\ell^-$ and of $\epe$ in a correlated manner while keeping $\varepsilon_K$ fully consistent with the SM expectations and its experimental value,
as seen in Figs.~2 and 3 of \cite{Aebischer:2023mbz}.

\begin{figure}
\centering%
\includegraphics[width=0.6\textwidth]{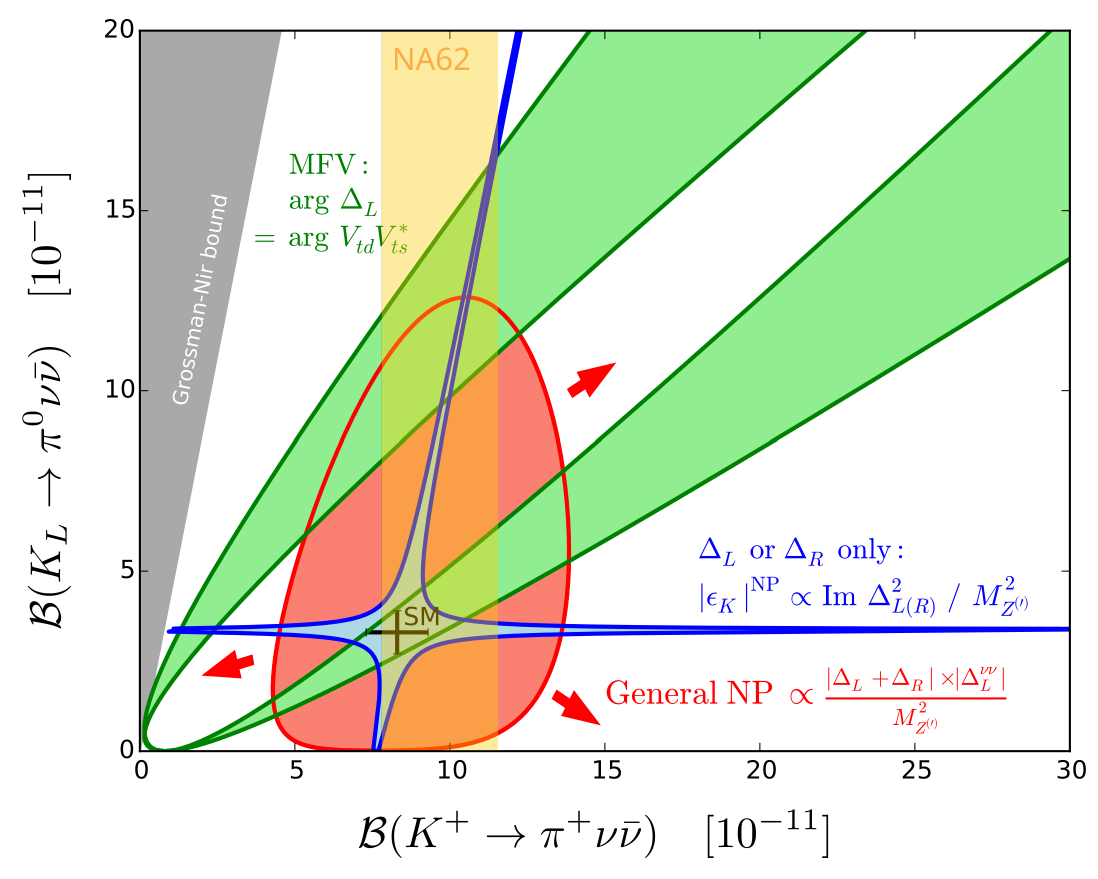}
\caption{\it Illustrations of common correlations in the $\mathcal{B}(\kpn)$ versus $\mathcal{B}(\klpn)$ plane. The expanding red region illustrates the lack of correlation for models with general LH and RH NP couplings. The green region shows the correlation present in models obeying Constrained Minimal Flavour Violation \cite{Buras:2000dm,Buras:2003jf,Blanke:2006ig}. The blue region shows the correlation induced by the constraint from $\varepsilon_K$ if only LH or RH couplings  are present \cite{Blanke:2009pq}.
The yellow band shows the values compatible with the NA62 measurement within $1\sigma$~\cite{NA62pnn2026}. Adapted from \cite{Buras:2015yca}.
}
\label{fig:illustrateEpsK}
\end{figure}

As seen in Fig.~\ref{fig:illustrateEpsK}, adapted from~\cite{Buras:2015yca} to include the new NA62 measurement, looking at $\mathcal{B}(\klpn)$ vs $\mathcal{B}(\kpn)$, the enhancements found in \cite{Aebischer:2023mbz} take place in the blue branch parallel to 
the Grossman-Nir bound \cite{Grossman:1997sk}. 
The type of correlation between $\mathcal{B}(\klpn)$ and $\mathcal{B}(\kpn)$ implied
by the $\varepsilon_K$ constraint in the presence of only LH or right-handed (RH) couplings 
had already been found in the context of the Littlest Higgs model with T-parity \cite{Blanke:2006eb} and in the minimal 3-3-1 model~\cite{Promberger:2007py}.
The general insight into such a correlation has been provided in~\cite{Blanke:2009pq}, and subsequently illustrated in an analysis of simplified models with LH and RH couplings \cite{Buras:2015yca}.

It is evident from Fig.~\ref{fig:illustrateEpsK} that the new NA62 measurement, together with the absence of NP in $\varepsilon_K$, implies a possible enhancement of the $\klpn$ rate by at most a factor of five in models with only LH or RH couplings, well below the model-independent Grossman-Nir bound~\cite{Grossman:1997sk}. 
If eventually $\kpn$ were proven to be essentially unaffected by NP, any significant  enhancements of $\mathcal{B}(\klpn)$, $\mathcal{B}(K_S\to\mu^+\mu^-)$ and $\epe$, as found in  \cite{Aebischer:2023mbz}, would be excluded in the case of only LH or RH couplings. The presence of both
LH and RH couplings is therefore required in order to populate regions outside the two blue branches.

In our view, the current experimental situation does not warrant a detailed sophisticated SMEFT analysis of the type presented in~\cite{Aebischer:2023mbz}.
Nevertheless, this will be of significant interest in the future, after the final result from the NA62 collaboration is published, the bounds for $\mathcal{B}(\klpn)$, $\mathcal{B}(K_S\to\mu^+\mu^-)$ improved and the situation of $\epe$ within the SM clarified.
The goal of our paper is, instead, to demonstrate in simple terms that indeed in the presence of both LH and RH $\bar s d$ couplings with new CP-violating phases $\klpn$, $K_S\to\mu^+\mu^-$, $K_L\to \pi^o\ell^+\ell^-$  and $\epe$ can be significantly enhanced while keeping both $\kpn$ and $\varepsilon_K$ SM-like.

Our paper is organised as follows. In Sec.~\ref{sec:2} we collect the most important formulae used in our paper.
In Sec.~\ref{sec:3} a specific $Z^\prime$ scenario is presented, for which the numerical analysis is performed in  Sec.~\ref{sec:4}. We end our paper with a short summary and an outlook for the coming years.

\section{General Formulae}\label{sec:2}

In this section we provide the general formulae for the branching ratios of the relevant kaon decays, both within and beyond the SM, where we allow for the presence of LH and RH quark currents. 
These expressions are sufficient to get an idea of the structure of various effects and to study them numerically. 
Further details, in particular the derivations of these formulae,
can be found in~\cite{Buras:2020xsm} and in the original papers listed there and below.

\boldmath
\subsection{$\kpn$ and $\klpn$} \label{sec:HeffRareKB}
\unboldmath

\subsubsection{Basic Expressions}

The branching ratios of the theoretically extremely clean $K\to\pi\nu\bar\nu$ decays can conveniently be parametrised in terms of a complex function $X_\text{eff}$ containing both SM and NP contributions.
Including isospin breaking corrections, summing over the three SM neutrino flavours\footnote{In this paper, we restrict our attention to the case of SM neutrinos in the final state.}, and generalising the SM formulae in~\cite{Buchalla:1996fp,Buchalla:1995vs,Buchalla:1998ba} to include NP effects from both LH and RH quark currents, one finds \cite{Buras:2015yca}
\begin{eqnarray}
\mathcal{B}(K^+\to\pi^+\nu\bar\nu)&=&\kappa_+ (1+\Delta_\text{EM})\cdot
\left[\left(\frac{{\rm Im}X_{\rm eff}}{\lambda^5}\right)^2+
\left(\frac{{\rm Re}\lambda_c}{\lambda}P_c(X)+
\frac{{\rm Re}X_{\rm eff}}{\lambda^5}\right)^2\right]\,,\label{bkpnn}\\
\mathcal{B}(\klpn)&=&\kappa_L\cdot
\left(\frac{{\rm Im}X_{\rm eff}}{\lambda^5}\right)^2,\label{bklpn}
\end{eqnarray}
\noindent
where $\lambda=\vus$ and $\lambda_c=V^*_{cs}V_{cd}$. Furthermore $\Delta_\text{EM}=-0.003$~\cite{Mescia:2007kn} parametrises the electromagnetic  corrections from photon exchanges, and the normalisation factors read~\cite{Mescia:2007kn}
\begin{equation}\label{kapp}
\kappa_+={ (5.173\pm 0.025 )\cdot 10^{-11}\left[\frac{\lambda}{0.225}\right]^8} , \quad 
\kappa_L=
(2.231\pm 0.013)\cdot 10^{-10}\left[\frac{\lambda}{0.225}\right]^8\,.
\end{equation}
$\kappa_+$ and $\kappa_L$ summarise all factors beyond CKM parameters, in particular the relevant hadronic matrix elements that can
be extracted from leading semi-leptonic decays of $K^+$, $K_L$ and $K_S$ mesons~\cite{Mescia:2007kn}.
In obtaining the numerical values in~(\ref{kapp}) 
\be\label{INPUT}
\sin^2\vartheta_W\equiv s_W^2=0.23116, \qquad \alpha(M_Z)=\frac{1}{127.9},
\ee
given in the $\overline{\text{MS}}$ scheme, have been used.
Their errors are below $0.1\%$ and can be neglected.
Further details can be found in Section 9.5.3 of \cite{Buras:2020xsm}.

Finally, the function $X_\text{eff}$ can be decomposed as follows:
\be\label{Xeff}
X_{\rm eff}=\lambda_t\,(X_L(K)+X_R(K))\equiv
\lambda_t X_{\rm tot}
\ee
where $\lambda_t = V_{ts}^*V_{td}$, and
\be
X_L(K)=X(x_t)+\Delta X_L(K)
\ee
represents contributions of LH currents, where $X(x_t)$ is the SM contribution with $x_t=m^2_t/M^2_W$ and $\Delta X_L(K)$ denotes NP contributions. 
Likewise $X_R(K)$ represents the contributions of RH currents that are absent in the SM. Note that only the {vector} parts of these currents contribute to the $K\to\pi\nu\bar\nu$ decays.

The NLO QCD corrections to $X(x_t)$ have been calculated in~\cite{Buchalla:1993bv,Misiak:1999yg,Buchalla:1998ba}, 
the NLO electroweak ones in~\cite{Brod:2010hi} and the NNLO QCD ones recently as stated below.
NLO QCD corrections to $P_c(X)$ have been calculated in~\cite{Buchalla:1993wq} 
and the NNLO ones in~\cite{Buras:2005gr,Buras:2006gb}, 
while the electroweak ones have been obtained in~\cite{Brod:2008ss}.
Finally the long-distance contributions to $P_c(X)$ have been determined in~\cite{Isidori:2005xm}.

The 2026 values of $X(x_t)$ and  $P_c(X)$ are then given by 
\be\label{PCNNLO}
X(x_t)=1.475\pm0.008, \qquad P_c(X)=0.405\pm 0.024\,,
\ee
with the latter calculated for $\lambda=0.225$. 
The NNLO value of $X(x_t)$ given above has been extracted in \cite{Buras:2026vbp} from~\cite{Stamou2025Kaon}. 
It replaces the value $X(x_t)=1.462\pm0.017$ presented in~\cite{Brod:2021hsj}.

Due to the absence of $P_c(X)$ in \eqref{bklpn},
$\mathcal{B}(K_L\to\pi^0\nu\bar\nu)$ has essentially no theoretical uncertainties. 
It is only affected by parametric uncertainties coming from $m_t$, $\imlt$ and $\kappa_L$, 
out of which only the one due to $\imlt$ is important.

Using the BV strategy of~\cite{Buras:2021nns,Buras:2022wpw}, that determines the CKM
parameters from $\Delta F=2$ observables and eliminates through ratios the dependence on $\vcb$,
the updated SM predictions of the $\kpn$ and $\klpn$ branching ratios determined independently of $\vcb$ and $\gamma$ read in 2026~\cite{Buras:2026vbp}
\begin{eqnarray}
\label{BV26}
\mathcal{B}(\kpn)_\text{SM} &=& (8.65\pm0.42)\times 10^{-11}\,,\\
\label{BV26L}
\mathcal{B}(\klpn)_\text{SM} &=& (3.05\pm 0.17)\times 10^{-11}\,
\end{eqnarray}
with the uncertainties in $\beta$, $\hat B_K$ and $P_c(X)$ taken into account.
These are the most precise determinations to date. 

The numbers in \eqref{BV26}, \eqref{BV26L} should be compared with the results of~\cite{Stamou2025Kaon} where instead of eliminating $\vcb$ a global UT fit of CKM parameters has been used:
\begin{eqnarray}\label{GSV}
\mathcal{B}(\kpn)_\text{SM} &=& (8.55\pm0.47)\times 10^{-11} \,,\\
\label{GSVL}
\mathcal{B}(\klpn)_\text{SM}&=&(3.15\pm 0.23)\times 10^{-11}\,.
\end{eqnarray}
One observes that there is a good agreement between these two ways of obtaining predictions for the $K\to\pi\nu\bar\nu$ branching ratios. 
This is not surprising since the global CKM fit is mostly based on $\Delta F=2$ observables.
Needless to say, the difference between the result in (\ref{BV26}) and (\ref{GSV}) is, 
from the point of view of the search for NP in $\kpn$,
immaterial as it is far beyond the precision of the present experimental data and the ones expected in the near future.
The significantly larger error in the $\klpn$ branching ratio in (\ref{GSVL})
is related to the $\vcb^4$ dependence present in $\mathcal{B}(\klpn)$ that is absent
in the BV strategy.

The latest NA62 measurement of the $\kpn$ branching ratio, quoted in (\ref{EXP19}), is fully consistent with the SM estimates, while still leaving room for some NP contributions.
For the $\klpn$ mode, the experimental sensitivity has not yet reached the level of the SM prediction, with an upper limit from the KOTO experiment quoted in (\ref{EXP19L}) leaving plenty of room for NP. 

\boldmath
\subsubsection{Interplay of $\kpn$ and $\klpn$}\label{Interplay}
\unboldmath
\label{sec:KpnnInterplay}

Once $\mathcal{B}(\kpn)$ and $\mathcal{B}(\klpn)$ have been measured, the
complex function $X_{\rm eff}$ can be determined, subject to
ambiguities that we will discuss below.

To this end it is useful to define
\be
X_{\rm eff}=-|X_{\rm eff}| e^{-i\beta_X}, \qquad
X_{\rm tot}=|X_{\rm tot}|e^{i\theta_X}, \qquad
\beta_X=\beta-\beta_s-\theta_X\,,
\ee
where
\be
V_{ts}= - \vts e^{-i\beta_s}, \qquad V_{td}=\vtd e^{-i\beta}\,.
\ee
The values of CKM parameters used in our analysis are collected in Appendix~\ref{CKM}.

As 
\be
   {\rm Re}\,X_{\rm eff} =-|X_{\rm eff}|\cos\beta_X, \qquad {\rm Im}\,X_{\rm eff} =|X_{\rm eff}|\sin\beta_X
\ee
equations (\ref{bkpnn}) and (\ref{bklpn}) can also be written as follows:
\begin{equation}\label{bkpnn2}
\mathcal{B}(K^+\to\pi^+\nu\bar\nu)=\kappa_+ (1+\Delta_\text{EM})\cdot
\left[\frac{|X_{\rm eff}|^2}{\lambda^{10}}+2\bar P_c(X)|X_{\rm eff}|
  \frac{\cos\beta_X}{\lambda^5} +\bar P_c(X)^2\right]\,,
\end{equation}
where we introduced
\be
\bar P_c(X)= - \frac{\RE \lambda_c}{\lambda} P_c(X) = \left(1-\frac{\lambda^2}{2}\right) P_c(X)\,,
\ee
and
\begin{equation}\label{bklpn2}
\mathcal{B}(\klpn)=\kappa_L\cdot
\frac{|X_{\rm eff}|^2}{\lambda^{10}}\sin^2\beta_X \,.
\end{equation}

Alternatively, we can solve (\ref{bkpnn}) and (\ref{bklpn}) for ${\rm Re}\,X_{\rm eff} $ and ${\rm Im}\,X_{\rm eff}$ \cite{Buras:2015yca}:
\begin{align}
{\rm Re}\,X_{\rm eff} &= \mp\lambda^5\left[\frac{\B(\kpn)}{\kappa_+(1 + \Delta_{\rm EM})} - \frac{\B(\klpn)}{\kappa_L}\right]^{1/2} - \lambda^4{\rm Re}\,\lambda_c P_c(X)\,,\label{ReX}\\
{\rm Im}\,X_{\rm eff} &= \pm\lambda^5\left[\frac{\B(\klpn)}{\kappa_L}\right]^{1/2}\,.\label{ImX}
\end{align}
These expressions can be used to directly determine $X_\text{eff}$ once the branching ratios of the decays $\kpn$  and $\klpn$ have been measured.
As evident from the signs $\mp$ and $\pm$ in front of the square-root terms,  a four-fold ambiguity remains in this determination. The upper sign in both formulae corresponds to the case in which NP contributions
 do not reverse the sign of the SM functions.
  As pointed out in \cite{Buras:2004ub}, within specific NP scenarios these ambiguities
can be resolved by considering other processes, such as the non-leptonic 
$B$ decays and the rare decays discussed in \cite{Buras:2004ub}.\footnote{Note that constraints from $B$ meson decays can only be meaningfully incorporated once a NP flavour structure is assumed. For instance \cite{Buras:2004ub} assumed a new flavour-universal contribution to the SM $Z^0$-penguin loop function $C(x_t)$.} Indeed
any process sensitive to electroweak penguins that contribute to
$X_{\rm eff}$ can be useful in this respect. This also applies to the ratio
$\epe$, as seen in the explicit expressions in \cite{Bosch:1999wr}. 

As pointed out in \cite{Buras:2004ub} it is possible to find a NP scenario
in which, despite a sizeable NP contribution to $X_\text{eff}$, $\mathcal{B}(\kpn)$ does not significantly
differ from the SM estimate because the enhancement of the first term in 
(\ref{bkpnn2}) can be  compensated by the suppression of the
second term. %($\cos \beta_X\ll\cos(\beta-\beta_s)$) and its reversed sign.
In this way, a spectacular enhancement of $\mathcal{B}(\klpn)$
by one order of magnitude can still be obtained.
Consequently,  while $\mathcal{B}(\klpn)\approx (1/3)\mathcal{B}(\kpn)$ 
in the SM, it can be substantially larger than 
$\mathcal{B}(\kpn)$ in such a NP  scenario.
As illustrated already in \cite{Buras:2004ub} and updated below,
the huge enhancement of $\mathcal{B}(\klpn)$ necessarily requires 
a large weak phase $\beta_X\approx 110^\circ$, as 
\be
\frac{\mathcal{B}(\klpn)}{\mathcal{B}(\klpn)_{\rm SM}}=
\left|\frac{X_{\rm tot}}{X_{\rm SM}}\right|^2
\left[\frac{\sin\beta_X}{\sin(\beta-\beta_s)}\right]^2\,, 
\ee
where values of $\beta$ and $\beta_{s}$ are given in Appendix~\ref{CKM}.
$\mathcal{B}(\klpn)$ can then be rather close to the Grossman-Nir bound~\cite{Grossman:1997sk}
\be\label{BOUND}
{\cal B}(\klpn)\le 4.3\, \mathcal{B}(\kpn).
\ee
This model-independent upper bound on ${\cal B}(\klpn)$ holds in any NP scenario with SM neutrinos in the final state. It could, however, be violated in models in which other massive dark particles appear in the final state, provided that their effect remains hidden from the NA62 analysis due to the  experimental setup~\cite{Fuyuto:2014cya}. Note that at the Grossman-Nir bound~\cite{Grossman:1997sk} the square root in (\ref{ReX}) vanishes.

\begin{figure}
\centering{\includegraphics[width=.6\textwidth]{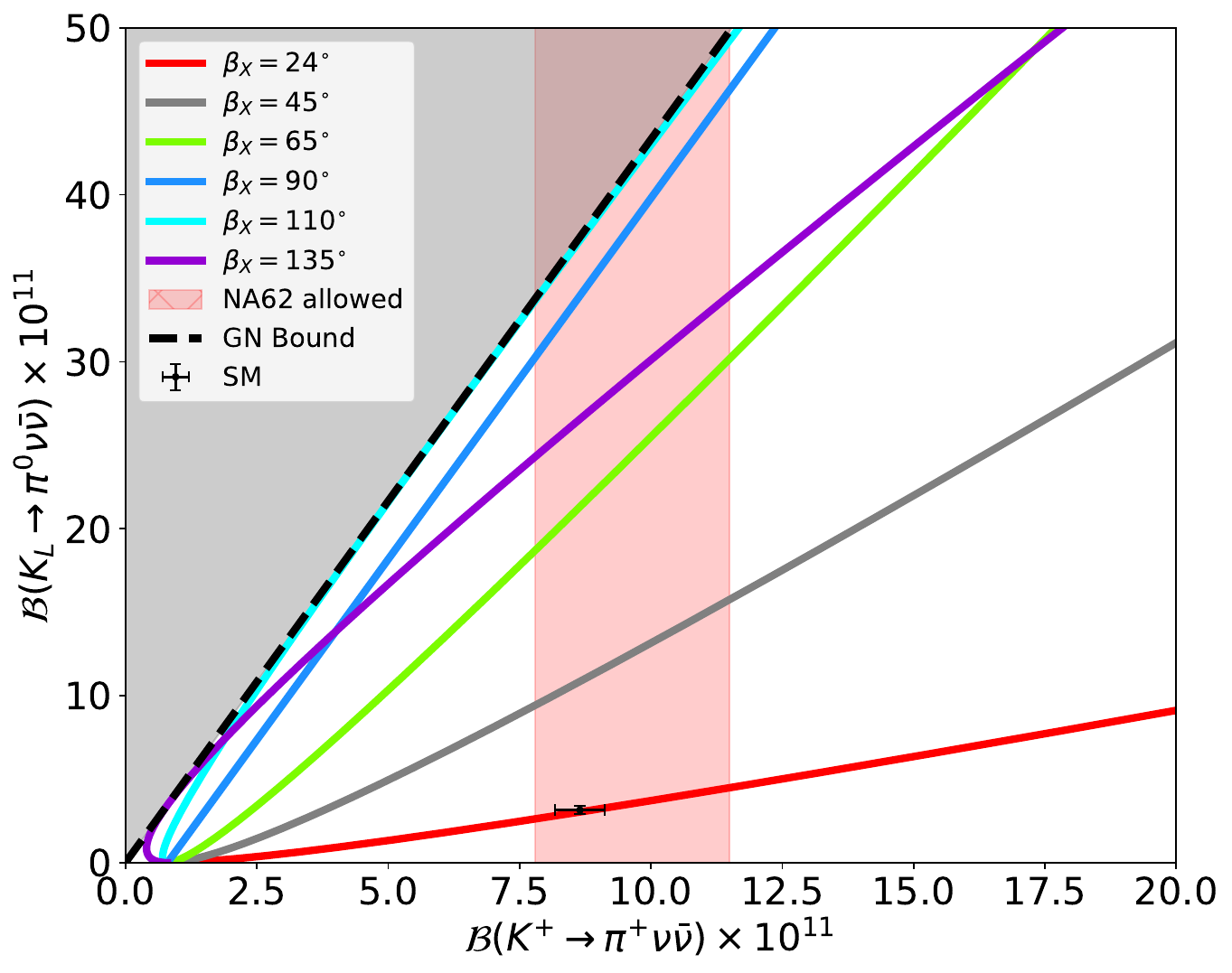}}
\caption{
Correlation between $\mathcal{B}(\kpn)$ and $\mathcal{B}(\klpn)$ for different values of the phase $\beta_X$.
}
\label{fig:betaX-Kpinunu}
\end{figure}

\begin{figure}
\centering{\includegraphics[width=.6\textwidth]{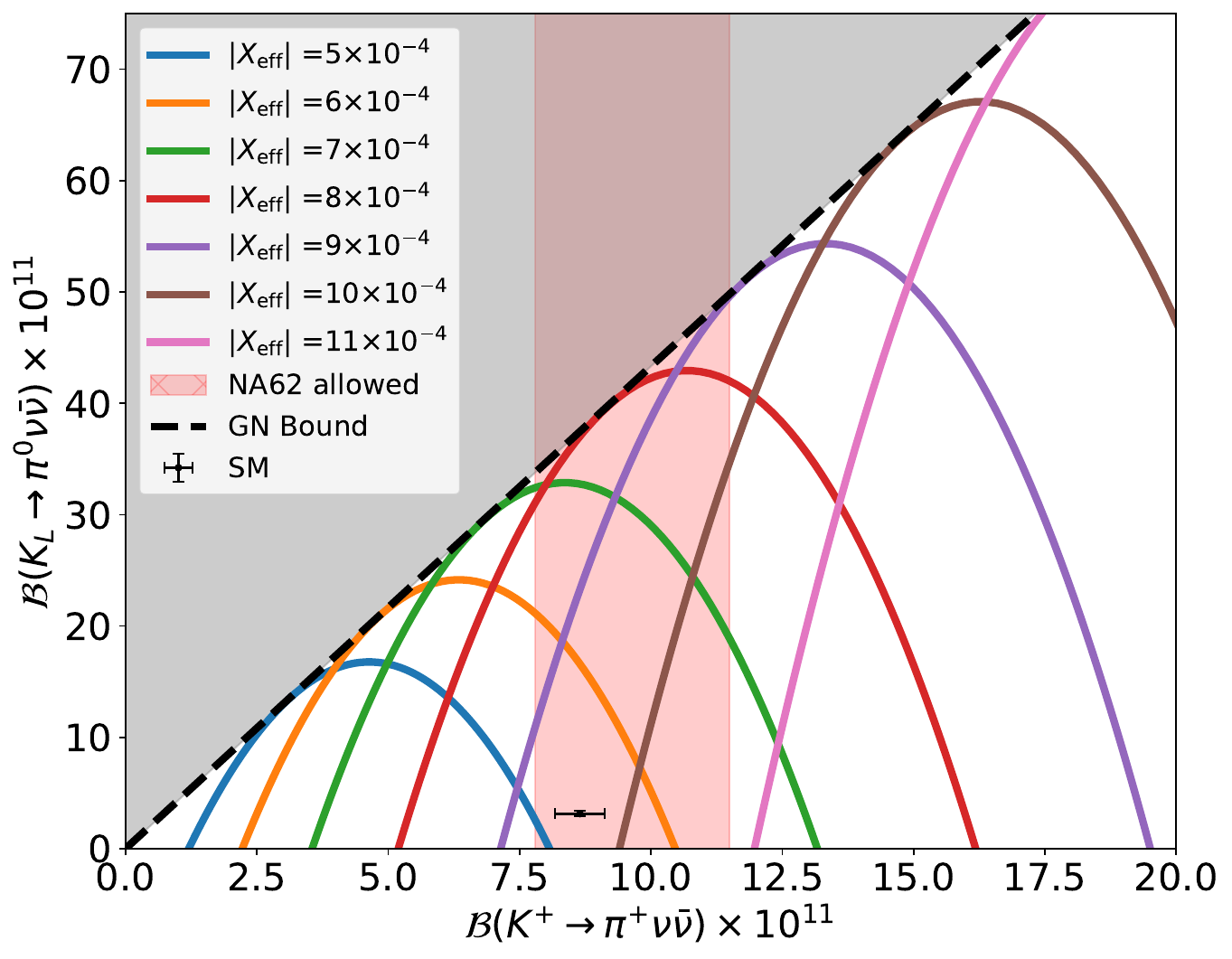}}
\caption{
Correlation between $\mathcal{B}(\kpn)$ and $\mathcal{B}(\klpn)$ for different values of $|X_\text{eff}|$.
}
\label{fig:Xeff-Kpinn}
\end{figure}

\begin{figure}
\centering{\includegraphics[width=.6\textwidth]{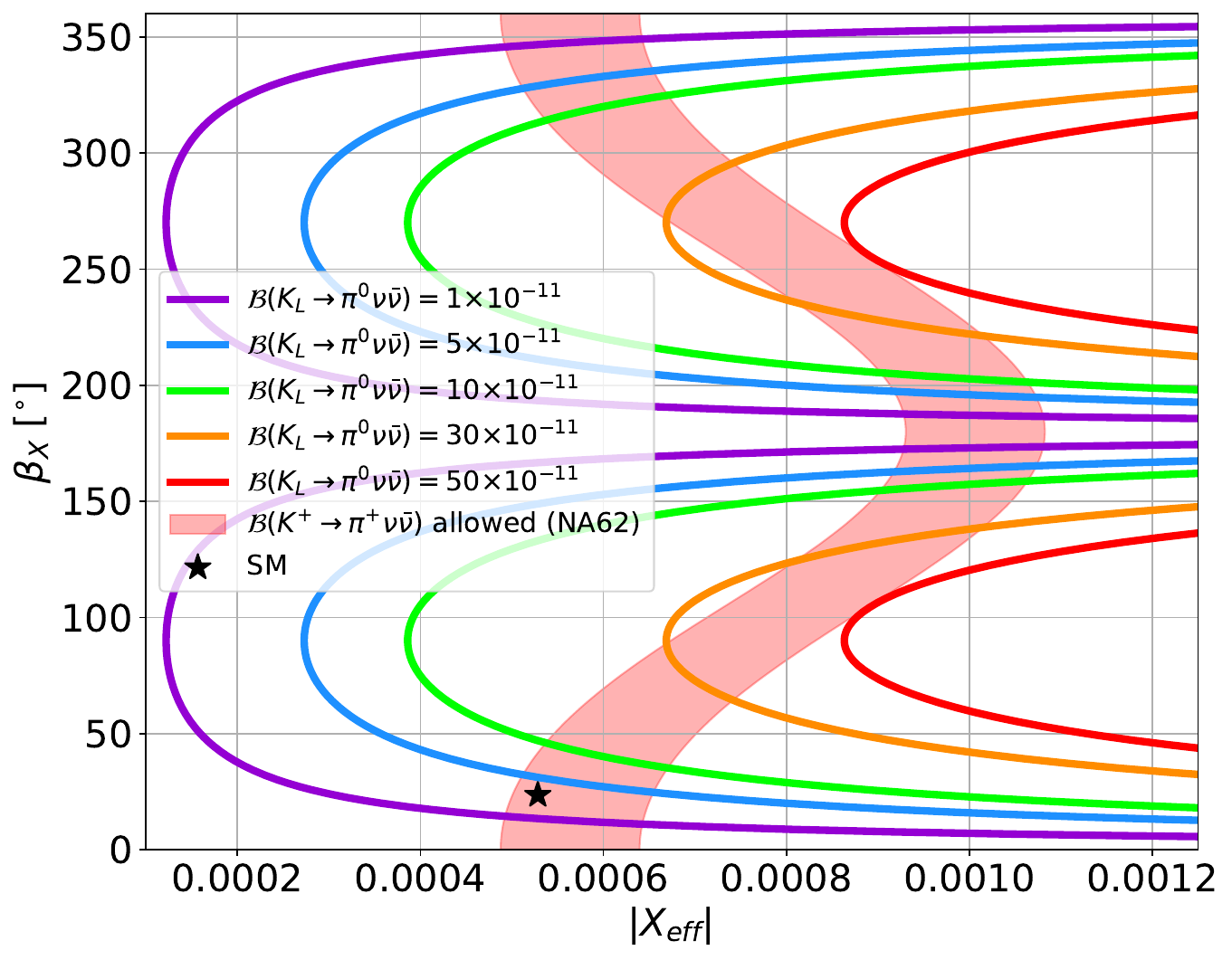}}
\caption{
Determination of $|X_\text{eff}|$ and $\beta_X$  from the $K\to \pi\nu\bar\nu$ decays. The red shaded region shows the $1\sigma$ allowed range for $\mathcal{B}(\kpn)$ by NA62, while the coloured lines display different hypothetical measurements of $\mathcal{B}(\klpn)$.
}
\label{fig:Xeff-det}
\end{figure}

Figs.\ \ref{fig:betaX-Kpinunu}--\ref{fig:Xeff-det} illustrate the impact of possible NP effects parametrised by the function $X_\text{eff}$ on the branching ratios of $\kpn$ and $\klpn$.
{In Fig.~\ref{fig:betaX-Kpinunu}, we show $\mathcal{B}(\klpn)$ as a function of $\mathcal{B}(\kpn)$ for fixed values of $\beta_X$. The magnitude of $|X_\text{eff}|$ is varied along the lines.
It is evident that the NA62 constraint, shown by the red band, implies that NP models with SM-like phases, $\beta_X \sim 24^\circ$, predict $\mathcal{B}(\klpn)$ to lie in the vicinity of its SM value, below $5\times 10^{-11}$.  

However, the prospects for $\mathcal{B}(\klpn)$ change drastically when allowing for a large complex phase $\beta_X$. 
In particular, for values in the ballpark of $\beta_X \sim 110^\circ$ the branching ratio of $\klpn$ is predicted close 
to the model-independent Grossman--Nir bound \cite{Grossman:1997sk} given in (\ref{BOUND}), shown by the black dashed line.
It is worth noting that for $\beta_X = 90^\circ$, shown in blue, the two branching ratios lie on a line parallel to the Grossman-Nir bound, with the offset determined by the charm contribution to $\kpn$.}

Analogously in Fig.~\ref{fig:Xeff-Kpinn}, we show $\mathcal{B}(\klpn)$ as a function of $\mathcal{B}(\kpn)$ for fixed values of $|X_\text{eff}|$, varying the phase $\beta_X$ along the different lines. We find that the NA62 measurement implies $|X_\text{eff}|$ to lie in the range 
\be
4.8\times 10^{-4} \lsim |X_\text{eff}| \lsim 10.9 \times 10^{-4}\,.
\ee
For larger (smaller) values of $|X_\text{eff}|$, too large (small) values for $\mathcal{B}(\kpn)$ are predicted irrespective of the value of $\beta_X$. Thus the current $\ord(20\%)$ experimental uncertainty by NA62 leads us to constrain $|X_\text{eff}|$ up to about a factor of two. Note that within the SM $|X_\text{eff}|$ reduces to $|\lambda_t X(x_t)| = 5.3 \times 10^{-4}$.

Fig.~\ref{fig:Xeff-det} shows how NA62 and KOTO II can determine $|X_{\rm eff}|$  and $\beta_X$:
the {red band indicates the $1 \sigma$ range from NA62, while} the coloured lines give the predicted $\klpn$ branching ratio as indicated by the legend. Again one clearly sees that {a large CP-violating phase $\beta_X$} is required for a large $\klpn$ branching ratio, while
keeping the $\kpn$ branching ratio consistent with the recent NA62 result.
It is worth noting that currently, without a measurement of ${\cal B}(\klpn)$, the value of $\beta_X$ is unconstrained in the model-independent consideration and it could take any value between 0 and $360^\circ$.

The four-fold ambiguity in the determination of 
$|X_{\rm eff}|$  and $\beta_X$ from the $K\to\pi\nu\bar\nu$ decays, already discussed below eqs.\ \eqref{ReX} and \eqref{ImX}, becomes evident from this figure. It is impossible to resolve it from these two decays alone. However, as we mentioned already, in specific NP scenarios other decays can help in resolving this ambiguity.

\boldmath
\subsection{$K_{L}\to \mu^+\mu^-$}\label{sec:KLmm}
\unboldmath
The decay $K_{L}\to \mu^+\mu^-$ is subject to significant long-distance (LD) contributions, preventing a precise prediction of its branching ratio.
Only the SD part of the dispersive contribution to $K_L\to\mu^+\mu^-$ can be reliably calculated. 
Despite this limitation, the constraint on the SD contribution places important bounds on certain NP scenarios.
In contrast to $\kpn$ and $\klpn$ (where only the vector current contributes) here the axial-vector current contributes.  
Relating the relevant matrix element $\langle 0|\bar s\gamma_\mu P_L d |K_L\rangle$ to the branching
ratio $\mathcal{B}(K^+\to\mu^+\nu_\mu)$ one finds~\cite{Buchalla:1993bv}
\be
{\mathcal{B}(K_L\to\mu^+\mu^-)_{\rm SD} = \kappa_\mu
\left( \frac{{\rm Re}\,Y^A_{\rm eff}}{\lambda^5} + \frac{{\rm Re}\,\lambda_c}{\lambda} P_c(Y)  \right)^2\,,}
\ee
where
\be
\kappa_\mu = 2.01\cdot 10^{-9}\,,\qquad  
Y(x_t)=(0.9882\pm 0.00024) Y_0(x_t)\,, \qquad P_c(Y) = 0.115\pm 0.017\,.
\ee
$Y(x_t)$ is the SM contribution at the NNLO level \cite{Bobeth:2013uxa}, where the LO Inami-Lim function is, to an excellent approximation, given by
\be
 Y_0(x_t) = 0.315 x_t^{0.78}\,,
 \ee
and 
$P_c(Y)$ represents the charm contribution at NNLO \cite{Gorbahn:2006bm}.

Similarly to $X_{\rm eff}$, we have introduced
\be\label{Yeff}
Y^A_{\rm eff} = V_{ts}^* V_{td} \,(Y(x_t)+\Delta Y_A(K))=-|Y^A_{\rm eff}|e^{-i\beta_Y}
\ee
and 
\be\label{YA}
Y^A_{\rm tot}=Y(x_t)+\Delta Y_A(K)=|Y^A_{\rm tot}|e^{i\theta_Y}
\ee
with $\beta_Y=\beta -\beta_s-\theta_Y$.
Here the NP contribution is represented by $\Delta Y_A(K)$.
For the $Z^\prime$ scenario it is given in (\ref{YLRK}).  

We find then
\be
\mathcal{B}(K_L\to\mu^+\mu^-)^\text{SM}_{\rm SD} = (8.48\pm0.73)\times10^{-10}\,.
\ee
The extraction of the SD part from data is subject to considerable uncertainties.
Here the important issue is the  sign of the interference of the SD dispersive
part $\chi_{\rm SD}$ of the decay amplitude of $K_L \to \mu^+\mu^-$ with the corresponding LD parts. 
Allowing for both possible signs implies a conservative bound $|\chi_{\rm SD}| \leq 3.1$~\cite{Isidori:2003ts}. 
This gives then the known upper bound~\cite{Isidori:2003ts}
\be\label{eq:KLmm-bound}
\mathcal{B}(K_L\to\mu^+\mu^-)_{\rm SD} \le 2.5 \times 10^{-9}\,,
\ee
roughly three times as large as the SM value.
This bound is also obtained for the sign favoured in
\cite{DAmbrosio:1996kjn,Gerard:2005yk} that implies $-1.7 \leq \chi_{\rm SD}\leq 3.1$.

On the other hand the opposite sign is favoured in~\cite{GomezDumm:1998gw}, 
giving  $-3.1 \leq \chi_{\rm SD}\leq 1.7$ and therefore approximately
\be\label{eq:KLmm-bound1}
\mathcal{B}(K_L\to\mu^+\mu^-)_{\rm SD} \le \mathcal{B}(K_L\to\mu^+\mu^-)^{\rm SM}_{\rm SD}\,.
\ee
It is well known that these bounds preclude large enhancements of $\mathcal{B}(\kpn)$ for models with NP governed by LH currents but are much less important if RH currents dominate the NP contributions.
Yet, in view of the recent NA62 result such an enhancement is no longer allowed anyway.

As already mentioned above it has been pointed out in~\cite{DAmbrosio:2017klp}
and further analysed in~\cite{DAmbrosio:2018ytt,DAmbrosio:2019xph,Dery:2021mct,Dery:2022yqc,DAmbrosio:2025mxa}
that a clean extraction of the SD parameters of the decay $K\to\mu^+\mu^-$ is possible from a measurement of the $K_L-K_S$ interference term in the time-dependent rate. 
Consequently direct CP violation can be measured in this decay.
This brings us to the next even more interesting decay.

\boldmath
\subsection{$K_{S}\to \mu^+\mu^-$}\label{sec:KSmm}
\unboldmath
The decay $\ksm$ provides a  sensitive probe of imaginary parts of new $\bar sd$ couplings. 
Its branching fraction receives LD and SD contributions, which are added incoherently in the total rate~\cite{Ecker:1991ru, Isidori:2003ts}. 
This is in contrast to the decay $\klm$, where LD and SD amplitudes interfere, 
and moreover $\BR(\klm)$ is sensitive to the real parts of $\bar sd$ couplings. 
The SD part of $\BR(\ksm)$ is given as
\begin{equation}
  \label{eq:ksm-br-SD}
 \BR(\ksm)_{\rm SD}
  = \tau_{K_S} \frac{G_F^2 \alpha^2}{8 \pi^3\sin^4\theta_W} m_K F_K^2 \sqrt{1-\frac{m^2_\mu}{m_K^2}} m_\mu^2\,
    \left[{\rm Im} Y^A_{\rm eff} \right]^2=1.04\times 10^{-5}\left[{\rm Im} Y^A_{\rm eff} \right]^2,
\end{equation}
with $ Y^A_{\rm eff}$ given in (\ref{Yeff}).

The SM prediction~\cite{Isidori:2003ts, DAmbrosio:2017klp} reads
\begin{align}\label{ISDA}
  \BR(\ksm)_{\rm SM} &
  = (4.99_{\rm LD} + 0.19_{\rm SD}) \times 10^{-12}
  = (5.2 \pm 1.5) \times 10^{-12},
\end{align}
about a factor of 50 below the present upper bound in (\ref{ksmbound}).
The predicted SD contribution therefore lies three orders of magnitude below the experimental bound.
As already mentioned previously it has been demonstrated in \cite{Dery:2021mct} that the short distance
contribution to $K_{S}\to\mu^+\mu^-$ can be extracted from data, making it another precision observable.

This is important because this decay---being dominated by direct CP violation in models with axial-vector currents---is very sensitive to NP contributions, as recently analysed in~\cite{Dery:2021vql}. 
For earlier studies see in particular the analyses of $Z^\prime$ models~\cite{Buras:2012jb}, 
leptoquark models~\cite{Bobeth:2017ecx}, and several models reviewed in~\cite{Buras:2020xsm}.

\boldmath
\subsection{$K_L\to\pi^0 \ell^+\ell^-$}\label{sec:KLpmm}
\unboldmath
The rare decays $K_L\to\pi^0e^+e^-$ and $K_L\to\pi^0\mu^+\mu^-$ are
dominated by CP-violating contributions.
KOTO II plans to have a second phase (after $K_L\to\pi^0\nu\bar\nu$) to measure $K_L\to\pi^0e^+e^-$ 
(the di-muon channel is also under investigation but is more challenging experimentally)~\cite{KOTO:2025gvq}.

In the SM the main contribution comes from the indirect (mixing-induced) CP violation and
its interference with the direct CP-violating contribution~\cite{D'Ambrosio:1998yj,Buchalla:2003sj,Isidori:2004rb,Friot:2004yr}.
The direct CP-violating contribution to the branching ratio is within the SM in the ballpark of $4\times 10^{-12}$, 
while the CP conserving contribution is at most $3\times 10^{-12}$. 
Among the rare $K$ meson decays, the decays in question are among the theoretically cleanest, 
but certainly cannot compete with the $K\to\pi\nu\bar\nu$ decays. 
Moreover, the dominant indirect CP-violating contributions are practically determined by the measured decays $K_S\to\pi^0\ell^+\ell^-$ and the parameter $\varepsilon_K$. 
Consequently they are not as sensitive as the $K_L\to\pi^0\nu\bar\nu$ decay to NP contributions, 
present only in the subleading direct CP violation. 
However, in the presence of large new CP-violating phases, 
the direct CP-violating contribution can become the dominant contribution and the branching ratios for $K_L\to\pi^0\ell^+\ell^-$ can be enhanced significantly, with a stronger effect in the case of $K_L\to\pi^0\mu^+\mu^-$ as already analysed in~\cite{Isidori:2004rb,Friot:2004yr,Mescia:2006jd}. Yet, even more important are the correlations of these decays with $\klpn$ and the ratio $\epe$ which we will encounter in Section~\ref{sec:4}.

In analogy to $X_{\rm eff}$ and $Y^A_{\rm eff}$, we have in these decays\footnote{Note that now $\bar\beta_Y$ and $\bar\theta_Y$ have to be introduced as the $K_{L}\to\pi^{0}\ell^{+}\ell^{-}$ decays are sensitive to the $\bar{s}d$ vector current.}
\begin{eqnarray}\label{YeffV}
Y^V_{\rm eff} &=& V_{ts}^* V_{td} \,(Y_0(x_t)+\Delta Y_V(K))=-|Y^V_{\rm eff}|e^{-i\bar\beta_Y}\,,\\
\label{YKV}
Y^V_{\rm tot} &=& Y_0(x_t)+\Delta Y_{V}(K)=|Y^V_{\rm tot}|e^{i\bar\theta_Y}\,,
\end{eqnarray}
and
\begin{eqnarray}\label{Zeff}
Z^V_{\rm eff} &=& V_{ts}^* V_{td} \,(Z_0(x_t)+\Delta Z_V(K))=- |Z^V_{\rm eff}|e^{-i\beta_Z} \,,\\
\label{ZK}
Z^V_{\rm tot} &= &Z_0(x_t)+\Delta Z_{V}(K)=|Z^V_{\rm tot}|e^{i\theta_Z}\,.
\end{eqnarray}

For the $Z^\prime$ scenario $\Delta Y_{V}(K)$ and $\Delta Z_{V}(K)$ are given in (\ref{ZLRK1}) and (\ref{ZLRK2}), respectively.
It should be noticed that in these decays even at the NLO level QCD corrections do not affect the SM one-loop functions $Y_0(x_t)$ and
$Z_0(x_t)$ which for $m_t = 162.69\,\text{GeV}$ are given by 
\be
Y_0(x_t)=0.315 x_t^{0.78},\qquad Z_0(x_t)= 0.175 x_t^{0.93}.
\ee
Only the $m_t$ independent contribution $P_0$ is affected by QCD and given at the NLO level by $P_0=2.88\pm 0.06$ \cite{Buras:1994qa}.

We have then~\cite{Isidori:2004rb,Friot:2004yr,Mescia:2006jd}
\be\label{eq:BrKpiLL}
\mathcal{B}(K_L\to\pi^0\ell^+\ell^-)=\left(C_\text{dir}^\ell\pm
  C_\text{int}^\ell\left|a_s\right| +
  C_\text{mix}^\ell\left|a_s\right|^2+C_\text{CPC}^\ell\right)
\times10^{-12}\,,
\ee
where
\begin{align}
&C_\text{dir}^e = (4.62\pm0.24)(\omega_{7V}^2+\omega_{7A}^2)\,,&\qquad&
C_\text{dir}^\mu =(1.09\pm0.05)(\omega_{7V}^2+2.32\omega_{7A}^2)\,,\\
&C_\text{int}^e = (11.3\pm0.3)\omega_{7V}\,,&\qquad&
C_\text{int}^\mu = (2.63\pm0.06)\omega_{7V}\,,\\
&C_\text{mix}^e = 14.5\pm0.05\,,&\qquad&
C_\text{mix}^\mu = 3.36\pm0.20\,,\\
&C_\text{CPC}^e \simeq 0\,,&\qquad&
C_\text{CPC}^\mu = 5.2\pm1.6\,,\\
&&&\hspace{-1.5cm}\left|a_s\right|=1.2\pm0.2
\end{align}
with
\bea
\omega_{7V} &=& \frac{1}{2\pi}\left[P_0+\frac{|Y^V_{\rm tot}|}{\sin^2\theta_W}
  \frac{\sin\bar\beta_Y}{\sin(\beta-\beta_s)}-4|Z^V_{\rm tot}|
  \frac{\sin\beta_Z}{\sin(\beta-\beta_s)}\right]\left[\frac{\IM
  \,\lambda_t}{1.4\cdot10^{-4}}\right]\,,\\
\omega_{7A} &=& -\frac{1}{2\pi}\frac{|Y^V_{\rm tot}|}{\sin^2\theta_W}
  \frac{\sin\bar\beta_Y}{\sin(\beta-\beta_s)}\left[\frac{\IM
  \,\lambda_t}{1.4\cdot10^{-4}}\right]\,,
\eea
where 
\be
\bar\beta_Y=\beta-\beta_s-\bar\theta_Y\,,\qquad 
\beta_Z=\beta-\beta_s-\theta_Z
\ee
with $Z^V_{\rm tot}$ defined in \eqref{ZK}.

The present experimental bounds
\be\label{KLLL}
\mathcal{B}(K_L\to\pi^0e^+e^-)<28\times 10^{-11}\quad\text{\cite{AlaviHarati:2003mr}}\,,\qquad
\mathcal{B}(K_L\to\pi^0\mu^+\mu^-)<38\times 10^{-11}\quad\text{\cite{AlaviHarati:2000hs}}
\ee
are still roughly one order of magnitude larger than the SM predictions,
which, using the CKM parameters of Appendix~\ref{CKM}, we find to be
\begin{align}
\mathcal{B}(K_L\to\pi^0e^+e^-)_\text{SM} &=
3.51\pm0.86 \left(1.56\pm0.54\right) \times10^{-11}\,,\label{eq:KLpee}
\\
{\mathcal{B}(K_L\to\pi^0\mu^+\mu^-)_\text{SM}} &= 
{1.40\pm0.26 \left(0.95\pm0.20\right) \times 10^{-11}}\,,\label{eq:KLpmm}
\end{align}
with the values in parentheses corresponding to the ``$-$'' sign in~\eqref{eq:BrKpiLL}. 
The central values of these predictions are very close to the original ones in \cite{Mescia:2006jd} but the uncertainties decreased. %visibly. \comMB{With my result ``decreased visibly'' is too strong, I would suggest ``decreased slightly''.}{\color{red}[Joel: perhaps we just say ``decreased''?]}

\boldmath
\subsection{$\epe$}\label{sec:epe}
\unboldmath

The parameter $\epe$ measures the amount of direct CP violation in $K\to\pi\pi$ decays. Its experimental world average, obtained by combining the results from the NA48~\cite{Batley:2002gn} and KTeV~\cite{AlaviHarati:2002ye, Worcester:2009qt} collaborations, reads
\be
\label{EXP}
    {(\epe)_\text{exp} 
    = (16.6 \pm 2.3) \times 10^{-4}\,.}
\ee

A short description of the present status of $\epe$ within the SM is given in \cite{Buras:2026vbp}. 
For our purposes it is sufficient to state that it is unclear at present whether the estimate of $\epe$ in the SM agrees with the result in (\ref{EXP}).

Indeed, the most recent result from lattice QCD, by the RBC-UKQCD collaboration~\cite{Abbott:2020hxn}, reads
\be
\label{RBCUKQCD}
  (\epe)_{\rm SM}^\text{LQCD} 
  = (21.7 \pm 8.4) \times 10^{-4} \,,
\ee
where statistical, parametric, and systematic uncertainties have been added in quadrature. 
On the other hand the most recent estimate of $\epe$ in the SM from ChPT~\cite{Cirigliano:2019ani,Gisbert:2020wkb} reads
\be
\label{Pich}
  (\epe)_\text{SM}^\text{ChPT}   = (14 \pm 5) \times 10^{-4} \,.
\ee
Both results are consistent with the experimental result in (\ref{EXP}),
but the large uncertainties still allow significant NP contributions.

Finally, based on the insight from Dual QCD~\cite{Buras:2015xba,Buras:2016fys} one finds~\cite{Buras:2020wyv}
\be
\label{AJBFINAL}
 (\epe)_{\rm SM}^\text{DQCD}= (5\pm2)\times 10^{-4}\,,
\ee
which, if true, would be one of the largest anomalies in flavour physics.

We will not repeat the arguments of the second author~\cite{Buras:2020wyv} why he thinks that the last result, 
even if rather uncertain, is closest to the true SM result among the three results quoted here. 
Instead we simply write
\be\label{GENERAL}
\frac{\varepsilon'}{\varepsilon}=\left(\frac{\varepsilon'}{\varepsilon}\right)^{\rm SM}+\left(\frac{\varepsilon'}{\varepsilon}\right)^{\rm NP}\,, \qquad 
\left(\frac{\varepsilon'}{\varepsilon}\right)^{\rm NP}= \kepe\cdot 10^{-3}, 
\ee
and, in view of the large uncertainties present in the SM estimates of $\epe$, we concentrate fully on the contribution of NP. 
In particular, we will investigate whether a positive shift in $\epe$ with $\kepe=\ord(1)$ can be generated by a $Z^\prime$ contribution.

\section[Specific $Z^\prime$ Scenario]{Specific $\bm{Z^\prime}$ Scenario}\label{sec:3}

\subsection{Preface}
An important class of models that can generate large NP contributions to rare meson decays are $Z^\prime$ models, 
in which the SM is accompanied by a new neutral vector boson related to a new $U(1)^\prime$ gauge symmetry. 
Within $Z^\prime$ models, the same flavour-changing quark couplings that enter $\Delta F=1$ flavour-changing neutral currents (FCNCs) also give rise to tree-level contributions to $\Delta F=2$ meson mixing observables.
As these can be measured with high precision, they impose stringent constraints on the parameter space of $Z^\prime$ models and limit their potential effects in FCNCs.

However, as emphasised in \cite{Buras:2014sba,Buras:2014zga,Crivellin:2015era}, the NP contributions to $\Delta F=2$ processes are suppressed for a particular pattern of left-handed~$\Delta_L^{q_1 q_2}(Z^\prime)$ and right-handed~$\Delta_R^{q_1 q_2}(Z^\prime)$ flavour-violating $Z^\prime$ couplings to quarks $q_1$ and $q_2$ defined through
\be
\vcenter{\hbox{\includegraphics[width=3cm]{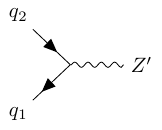}}}\qquad ig_\mu^{q_1q_2}(Z')=i\gamma_\mu [\Delta_L^{q_1 q_2}(Z')P_L+\Delta_R^{q_1 q_2}(Z')P_R].
\ee

As stressed in particular in \cite{Buras:2014sba}, in the case of the decays considered by us this implies that the $\Delta_R^{sd}(Z^\prime)$ coupling must be roughly two orders of magnitude smaller than $\Delta_L^{sd}(Z^\prime)$ in order to keep $\varepsilon_K$ consistent with experiment. 
We will be more explicit about this in the next section.
As we shall see, this simplifies our analysis significantly while allowing 
$\Delta_L^{sd}(Z^\prime)$ to be sufficiently large to provide significant NP
contributions to the rare kaon decays and the ratio $\epe$ considered by us.

A compendium of the $Z^\prime$ contributions to the observables considered by us has been presented in~\cite{Buras:2012jb} 
and can also be found in Section 15.4 of \cite{Buras:2020xsm}. 
From these references, we extract the expressions used in the present paper. 
Note that we still keep $\Delta_R^{sd}(Z^\prime)$ in these expressions in order to make explicit the simplifications resulting from its smallness.

\boldmath
\subsection{$\varepsilon_K$ and $\Delta M_K$}\label{EK}
\unboldmath

\begin{figure}
\centering{
\includegraphics[width=.5\textwidth]{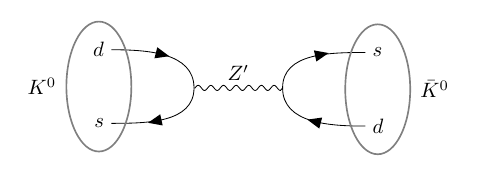}
}
\caption{Feynman diagram for $K^0-\bar K^0$ mixing in $Z'$ models.\label{feyn:KKbar}}
\end{figure}

The contribution of a $Z^\prime$ boson to the parameter $\varepsilon_K$, measuring CP violation in $K^0-\bar K^0$ mixing, and the mass difference $\Delta M_K$ are depicted in Fig.\ \ref{feyn:KKbar} and given as follows
\be
\varepsilon_K(Z^\prime)=\frac{\kappa_\varepsilon e^{i\phi_\varepsilon}}{\sqrt{2}(\Delta M_K)_{\rm exp}}
\text{Im}\left((M_{12})_{Z^\prime}^{sd}\right), \qquad \Delta M_K(Z^\prime)= 2\text{Re}\left((M_{12})_{Z^\prime}^{sd}\right),
\ee
where $\kappa_\epsilon = 0.94(2)$~\cite{Buras:2008nn,Buras:2010pza}.
They should simply be added to the SM contributions.

In the presence of both LH and RH couplings of a $Z^\prime$ gauge boson to SM quarks, the left-right (LR) $\Delta F=2$ operators are generated, whose contributions to the mixing amplitude $M_{12}^{sd}$ are enhanced through renormalisation group effects relative to left-left (VLL) and right-right (VRR) operators. 
Moreover, in the case of $M_{12}^{sd}$ additional chiral enhancements of the hadronic matrix elements of LR operators are present. 
As pointed out in~\cite{Buras:2014sba} this fact can be used to suppress NP contributions to $\varepsilon_K$ and $\Delta M_K$ through some fine-tuning between VLL, VRR and LR contributions, thereby allowing for larger contributions to other observables.

This idea, used in~\cite{Buras:2014zga}, is also very useful for our present analysis.
To this end, one writes the $Z^\prime$ contributions to the mixing amplitude as follows~\cite{Buras:2012jb}: 
\be\label{ZpnewK}
(M_{12}^*)_{Z^\prime}^{sd}  =  \frac{\left(\Delta_L^{sd}(Z')\right)^2}{2M_{Z^\prime}^2} \langle \hat Q_1^\text{VLL}(M_{Z^\prime})\rangle^{sd} z_{sd},
\ee
where $z_{sd}$ is complex and given by
\be\label{deltasupp}
 z_{sd}=\left[1+\left(\frac{\Delta_R^{sd}(Z')}{\Delta_L^{sd}(Z')}\right)^2+2\kappa_{sd}\frac{\Delta_R^{sd}(Z')}{\Delta_L^{sd}{Z')}}\right],
\qquad \kappa_{sd}=\frac{\langle \hat Q_1^\text{LR}(M_{Z^\prime})\rangle^{sd}}{\langle \hat Q_1^\text{VLL}(M_{Z^\prime})\rangle^{sd}}.
\ee

Following \cite{Buras:2014zga}, and using the technology of \cite{Buras:2001ra,Buras:2012fs}, we have expressed $z_{sd}$ in terms of the renormalisation-scheme-independent matrix elements
$\langle\hat Q_1^\text{VLL}(M_{Z^\prime})\rangle^{sd}$ and $\langle \hat Q_1^\text{LR}(M_{Z^\prime})\rangle^{sd}$, 
which are defined in eqs. (35) and (36) of \cite{Buras:2014zga}. 
Table 5 of \cite{Buras:2014zga} lists the values of these matrix elements for different $M_{Z^\prime}$. 
For $M_{Z^\prime}=5\tev-20\tev$ one finds $\kappa_{sd}(M_{Z^\prime})\approx -120 $, so that the requirement $z_{sd}\approx 0$ implies 
\be\label{240}
\frac{\Delta_R^{sd}(Z')}{\Delta_L^{sd}(Z')} \approx \frac{1}{240}.
\ee
We thus conclude that, in order to suppress $Z'$ contributions to $K^0-\bar K^0$ mixing by means of eq.\ \eqref{240}, 
the only role of the $\Delta_R^{sd}(Z')$ coupling is to keep $\varepsilon_K$ and $\Delta M_{K}$ SM-like without any impact on other observables. 
This in turn simplifies our numerical analysis, as will be evident in the formulae below.

\boldmath
\subsection{$\kpn$ and $\klpn$} \label{sec:bfg1}
\unboldmath

\begin{figure}
\centering{
\includegraphics[width=.4\textwidth]{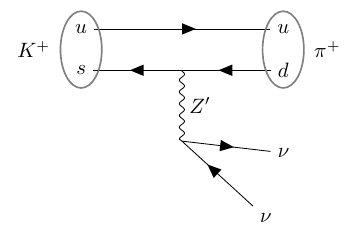}
\hspace{1cm}
\includegraphics[width=.4\textwidth]{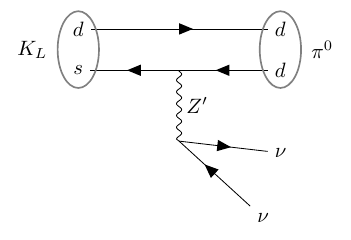}
}
\caption{Feynman diagrams for $\kpn$ (left) and $\klpn$ (right) in $Z'$ models.\label{feyn:Kpnn}}
\end{figure}

The Feynman diagrams for a $Z'$ exchange contributing to $\kpn$ and $\klpn$ can be found in Fig.\ \ref{feyn:Kpnn}. They imply contributions to the expressions collected in Section \ref{sec:HeffRareKB} as follows:
\be\label{XLRK}
\Delta X_{L}(K)=\frac{\Delta_L^{\nu\bar\nu}(Z')}{g^2_{\rm SM}M_{Z'}^2} 
                            \frac{\Delta_L^{sd}(Z')}{V_{ts}^* V_{td}}\,,
\qquad
X_{R}(K)=\frac{\Delta_L^{\nu\bar\nu}(Z')}{g^2_{\rm SM}M_{Z'}^2}
                            \frac{\Delta_R^{sd}(Z')}{V_{ts}^* V_{td}} % \approx 0,
\ee
and therefore
\begin{equation}
    X_{\rm eff} = \lambda_{t}X(x_{t}) + 
    \frac{1}{g_\text{SM}^{2}M_{Z^{\prime}}^{2}}\Delta_{L}^{\nu\bar{\nu}}
    \left( \Delta_{L}^{sd}(Z^{\prime}) + \Delta_{R}^{sd}(Z^{\prime}) \right) \,. 
\end{equation}
We further introduced 
\be\label{gsm}
g_{\text{SM}}^2=4\frac{G_F}{\sqrt 2}\frac{\alpha}{2\pi\sin^2\theta_W}=1.78137\times 10^{-7} \gev^{-2}\,.
\ee
In what follows we neglect the small $\Delta_R^{sd}(Z^\prime)$ contribution  in view of (\ref{240}), so that $X_{R}(K)=0$.

\boldmath
\subsection{$K_{L}\to \mu^+\mu^-$, $K_{S}\to \mu^+\mu^-$ and $K_L\to\pi^0\ell^+\ell^-$}\label{KSmmZp}
\unboldmath

\begin{figure}
\centering{
\includegraphics[width=.45\textwidth]{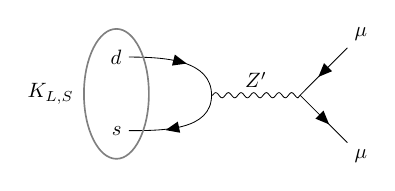}
\hspace{1cm}
\includegraphics[width=.4\textwidth]{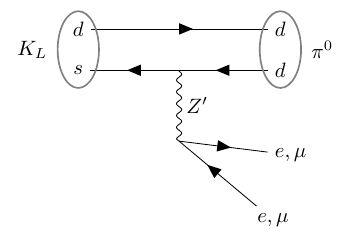}
}
\caption{Feynman diagrams for $K_{L,S}\to\mu^+\mu^-$ (left) and $K_L\to\pi^0\ell^+\ell^-$ (right) in $Z'$ models.\label{feyn:Kll}}
\end{figure}

The Feynman diagrams for a $Z'$ exchange contributing to {the decays} $K_{L,S}\to\mu^+\mu^-$ and $K_L\to\pi^0\ell^+\ell^-$ can be found in 
Fig.\ \ref{feyn:Kll}.

For the leptonic modes $K_{L}\to \mu^+\mu^-$ and $K_{S}\to \mu^+\mu^-$ the relevant contribution to the expressions collected in Sections \ref{sec:KLmm} and \ref{sec:KSmm}
is given by
\be\label{YLRK}
\Delta Y_{A}(K)=\frac{\Delta_A^{\mu\bar\mu}(Z')}{g^2_{\rm SM}M_{Z'}^2}
\frac{(\Delta_L^{sd}(Z')-\Delta_R^{sd}(Z'))}{V_{ts}^* V_{td}}
\ee
and therefore
\begin{equation}
    Y^{A}_{\rm eff} = \lambda_{t}Y_{0} + 
    \frac{1}{g_\text{SM}^{2}M_{Z^{\prime}}^{2}}\Delta_{A}^{\mu\bar{\mu}}
    \left( \Delta_{L}^{sd}(Z^{\prime}) - \Delta_{R}^{sd}(Z^{\prime}) \right) \,,
\end{equation}
where we can again set $\Delta_R^{sd}(Z^\prime)\approx 0$ in view of (\ref{240}).

For the semileptonic $K_L\to\pi^0\ell^+\ell^-$ modes discussed in Section~\ref{sec:KLpmm} one finds~\cite{Blanke:2006eb,Blanke:2008yr,Buras:2012jb}
\be\label{ZLRK1}
\Delta Y_{\rm V}(K)=\frac{\Delta_A^{\mu\bar\mu}(Z')}{g^2_{\rm SM}M_{Z^\prime}^2}
\frac{\Delta_V^{sd}(Z')}{V_{ts}^* V_{td}} %\approx \frac{\Delta_A^{\mu\bar\mu}(Z')}{g^2_{\rm SM}M_{Z^\prime}^2}
%\frac{\Delta_L^{sd}(Z')}{V_{ts}^* V_{td}}
\ee
and
\be\label{ZLRK2}
\Delta Z_{V}(K)=\frac{1}{4\sin^2\theta_W}\frac{2\Delta_R^{\mu\bar\mu}(Z^\prime)}{g^2_{\rm SM}M_{Z'}^2} \frac{\Delta_V^{sd}(Z^\prime)}{V_{ts}^* V_{td}}
\ee
with
\be
\Delta_A^{\mu\bar\mu}(Z')=\Delta_R^{\mu\bar\mu}(Z')-\Delta_L^{\mu\bar\mu}(Z'),
\qquad 
\Delta_V^{sd}(Z')=\Delta_R^{sd}(Z')+\Delta_L^{sd}(Z')\approx \Delta_L^{sd}(Z') .
\ee
Again in view of \eqref{240} we neglected the small contributions from $\Delta_R^{sd}(Z')$.

\boldmath
\subsection{$\epe$}
\unboldmath

\begin{figure}
\centering{
\includegraphics[width=.4\textwidth]{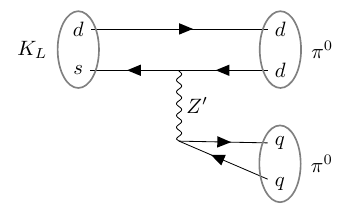}
}
\caption{Representative Feynman diagram for $\epe$ in the $Z'$ scenario.}
\label{feyn:epsprime}
\end{figure}

Due to the large
uncertainties present in the SM estimates of $\epe$, discussed in Section~\ref{sec:epe}, 
we concentrate on the NP contribution characterised by the parameter $\kepe$ defined in (\ref{GENERAL}). 
As demonstrated in \cite{Buras:2015jaq}, 
the NP contribution of $Z^\prime$ models is dominated by the QCD penguin operator $Q_6$ and the electroweak penguin operator $Q_8$. 
However, as demonstrated in~\cite{Aebischer:2020mkv}, 
the $Q_6$ scenario is ruled out by significant contributions to $\varepsilon_K$ implied by renormalisation-group effects. 
The latter is not the case in the $Q_8$ scenario. 
Consequently, the NP contribution to $\epe$ in our $Z^\prime$ scenario is dominated by the electroweak penguin operator
\be
Q_8 = \frac{3}{2}\,(\bar s_{\alpha} d_{\beta})_{V-A}\!\!\sum_{q=u,d,s,c,b,t}
      e_q\,(\bar q_{\beta} q_{\alpha})_{V+A} \,.
      \ee
Within $10\%$ precision it can be written  as follows~\cite{Buras:2015jaq}:
\be\label{eprimeZprimeQ8}
\left(\frac{\varepsilon'}{\varepsilon}\right)^{\rm NP}=\left(\frac{\varepsilon'}{\varepsilon}\right)^L_{Z^\prime}= 14.5 \,\bei \,\left[\frac{5\tev}{M_{Z^\prime}}\right]^2 {\IM}\left(\Delta_L^{s d}(Z^\prime)\right)\Delta_R^{q \bar q}(Z^\prime)\,.
\ee
This accuracy is sufficient for our analysis, 
not only because of the non-perturbative uncertainties in the SM contribution, 
but also because of the new coupling parameter $\Delta_R^{q \bar q}(Z^\prime)$ entering the game. 
In view of the smallness of the coupling $\Delta_R^{s d}(Z^\prime)$, see \eqref{240},
the contribution of $Q^\prime_8$ is also negligible. 
A representative Feynman diagram is shown in Fig.\ \ref{feyn:epsprime}.

Note that in writing \eqref{eprimeZprimeQ8} we have introduced the real coupling parameter $\Delta_R^{q \bar q}(Z^\prime)$ as the effective flavour-conserving $Z'$ coupling to RH quarks entering the electroweak penguin contribution to $\epe$, while ensuring the dangerous QCD penguin contribution to be small. See Section 4 of \cite{Aebischer:2020mkv} for details.
    
In what follows we will set the non-perturbative factor
$\bei=0.80$ as this value is consistent with both the Dual QCD approach~\cite{Buras:2015xba} 
and the lattice QCD result by the RBC-UKQCD collaboration~\cite{Abbott:2020hxn}.

It should be remembered that, 
although there is considerable freedom in the values of the diagonal quark and lepton couplings of $Z^\prime$, 
one must ensure that they are consistent with the relevant LEP II and LHC bounds.
Concerning the latter, the study in~\cite{deVries:2014apa} implies
\be \label{LHCbound}
\left|\Delta_R^{q \bar q}(Z^\prime) \right| \leq 1.7 \left[ \frac{M_{Z^\prime}}{5\tev}
 \right] \left[ 1 + \left(\frac{1.3\tev}{M_{Z^\prime}}\right)^2\right]\,.
\ee
This constraint is easily satisfied by the values of $\Delta_R^{q \bar q}(Z^\prime)$ used in our numerical analysis.
On the other hand, bounds on the leptonic $Z^\prime$ couplings can be extracted from the final analysis of the LEP-II data~\cite{Schael:2013ita}.
Yet, there is still considerable freedom, as those bounds are for products of electron and other lepton couplings. 
Still, to respect $\text{SU(2)}_L$ symmetry, one has to ensure that
 $\Delta^{\nu\bar\nu}_L(Z^\prime)=\Delta^{e\bar e}_L(Z^\prime)$.
Furthermore, lepton universality implies $\Delta^{e\bar e}_{L,R}(Z^\prime) = \Delta^{\mu\bar\mu}_{L,R}(Z^\prime)$.

\section[Numerical Analysis for {$Z^\prime$} Scenario]{Numerical Analysis for $\bm{Z^\prime}$ Scenario}\label{sec:4}
\sectionmark{Numerical Analysis for $Z^\prime$ Scenario}

The simple $Z^\prime$ scenario considered in the present paper is described by the following
\begin{itemize}
\item
  four real parameters
  \be
  \Delta_L^{\nu\bar\nu}(Z^\prime) = \Delta_L^{\ell\bar\ell}(Z^\prime), \qquad \Delta_R^{\ell\bar\ell}(Z^\prime),\qquad \Delta^{q\bar{q}}_R(Z^\prime),\qquad M_{Z^\prime} ,
  \ee
\item
  and two complex parameters
  \be
  \Delta_L^{sd}(Z^\prime), \qquad  \Delta_R^{sd}(Z^\prime), 
  \ee 
\end{itemize}
which amounts to eight real parameters in total. Note that we have assumed $Z'$ to be a $U(1)'$ gauge boson with lepton-flavour-universal couplings.

For our numerical analysis we choose $M_{Z^{\prime}}=5\,\text{TeV}$ and, from~(\ref{240}) that demonstrates $\Delta_{R}^{sd}(Z^{\prime})\ll\Delta_{L}^{sd}(Z^{\prime})$, we set $\Delta_{R}^{sd}(Z^{\prime})=0$. 
This reduces our parameter space to three real and one complex parameter(s),
\begin{equation}\label{eq:param}
    \Delta_L^{\ell\bar\ell}(Z^\prime), \qquad \Delta_R^{\ell\bar\ell}(Z^\prime) , \qquad \Delta^{\bar q q}_R(Z^\prime), \qquad \Delta_L^{sd}(Z^\prime), 
\end{equation}
leaving us with a five-dimensional parameter space to explore.

In our numerical analysis we investigate the following quantities:
\begin{eqnarray}
R_{\nu\bar\nu}^+  &= & \frac{\mathcal{B}(K^+ \to \pi^+ \nu \bar \nu)}{\mathcal{B}(K^+ \to \pi^+ \nu \bar \nu)_\text{SM}} 
\,, \quad
R_{\nu\bar\nu}^0 =  \frac{\mathcal{B}(K_L \to \pi^0 \nu \bar \nu)}{\mathcal{B}(K_L \to \pi^0 \nu \bar \nu)_\text{SM}}
\,, \quad 
R_{\mu^+\mu^-}^L =\frac{\mathcal{B}(K_L \to \mu^+\mu^-)_{\rm SD}}{\mathcal{B}(K_L\to \mu^+\mu^-)^\text{SD}_\text{SM}}
\,, \nonumber\\
R_{\mu^+\mu^-}^S  &=&  \frac{\mathcal{B}(K_S \to \mu^+\mu^-)_{\rm SD}}{\mathcal{B}(K_S \to \mu^+\mu^-)^\text{SD}_\text{SM}} 
\,,\quad 
R^{0}_{\pi\ell^{+}\ell^{-}} = \frac{\mathcal{B}(K_L \to \pi^0 \ell^+ \ell^-)}{\mathcal{B}(K_L \to \pi^0 \ell^+ \ell^-)_\text{SM}}
\,,\quad
\kappa_{\varepsilon^{\prime}} = 10^{3} \times \left(\frac{\varepsilon'}{\varepsilon}\right)^{\rm NP}\,.
\label{eq:kappas}
\end{eqnarray}
The dependencies of these quantities of interest, given in equation~\eqref{eq:kappas}, on the parameters listed in \eqref{eq:param} are summarised in Table~\ref{tab:VariableDepenciesRefactored}.

\begin{table}
    \centering
    \begin{tabular}{|c||c|c|c||c|c|}
        \hline
        Variable
            & $\Delta_{L}^{\ell\bar{\ell}}(Z^{\prime})$ 
            & $\Delta_{R}^{\ell\bar{\ell}}(Z^{\prime})$ 
            & $\Delta_{R}^{q\bar{q}}(Z^{\prime})$ 
            & $\text{Re}\left(\Delta_{L}^{sd}(Z^{\prime})\right)$
            & $\text{Im}\left(\Delta_{L}^{sd}(Z^{\prime})\right)$
        \\
        \hline
        $R_{\nu\bar{\nu}}^{+}$ 
            & \checkmark & - & - & \checkmark & \checkmark \\
        $R_{\nu\bar{\nu}}^{0}$ 
            & \checkmark & - & - & - & \checkmark \\
        \hline
        $R_{\mu^+{\mu}^-}^{L}$ 
            & \checkmark & \checkmark & - & \checkmark & - \\
        $R_{\mu^+{\mu^-}}^{S}$ 
            & \checkmark & \checkmark & - & - & \checkmark \\
        \hline
        $R^{0}_{\pi\ell^{+}\ell^{-}}$
            & \checkmark & \checkmark & - & - & \checkmark \\ 
        \hline
        $\kappa_{\varepsilon^{\prime}}$
            & - & - & \checkmark & - & \checkmark \\
        \hline
    \end{tabular}
    \caption{Functional dependencies of each variable, assuming a fixed value of $M_{Z}^{\prime}=5\,\text{TeV}$, $\Delta_{R}^{sd}(Z^{\prime})\approx 0$ and flavour-universal couplings to leptons.}
    \label{tab:VariableDepenciesRefactored}
\end{table}

With this explicit setup, we want to demonstrate that the following set of requirements can be satisfied simultaneously:
\begin{enumerate}
\item
    The $K^0-\bar K^0$ mixing parameters $\varepsilon_K$ and $\Delta M_K$ are SM-like.
\item 
    $\mathcal{B}(K^{+}\rightarrow\pi^{+}\nu\bar{\nu})$ is consistent with the SM, and within the experimentally-allowed range from the latest NA62 measurement.
\item 
    $\mathcal{B}(K_{L}\rightarrow\pi^{0}\nu\bar{\nu})$, to be measured by KOTO II, can be significantly enhanced by a factor of $\mathcal{O}(10)$ with respect to its SM value.
\item 
    The upper bound on {the SD contribution} to $K_L\to \mu^+\mu^-$ can be satisfied.
\item
    $\mathcal{B}(K_S\to \mu^+\mu^-)$, searched for by LHCb, can be significantly enhanced with respect to its SM value.
\item
    The branching ratios of the decays  $K_L\to \pi^0\ell^+\ell^-$ can be significantly enhanced, while satisfying the experimental upper limits.
\item
    $\epe$ can be significantly enhanced with respect to its SM value, with $\kappa_{\varepsilon^{\prime}}=\mathcal{O}(1)$.
\end{enumerate}

Point 1 on this list is satisfied by our choice
$\Delta_{R}^{sd}(Z^{\prime}) \approx (1/240) \Delta_{L}^{sd}(Z^{\prime})\approx 0$ which ensures $\varepsilon_K$ and $\Delta M_K$ are SM-like, 
regardless of the values of the other parameters.

\begin{figure}
    \includegraphics[width=0.51\linewidth]{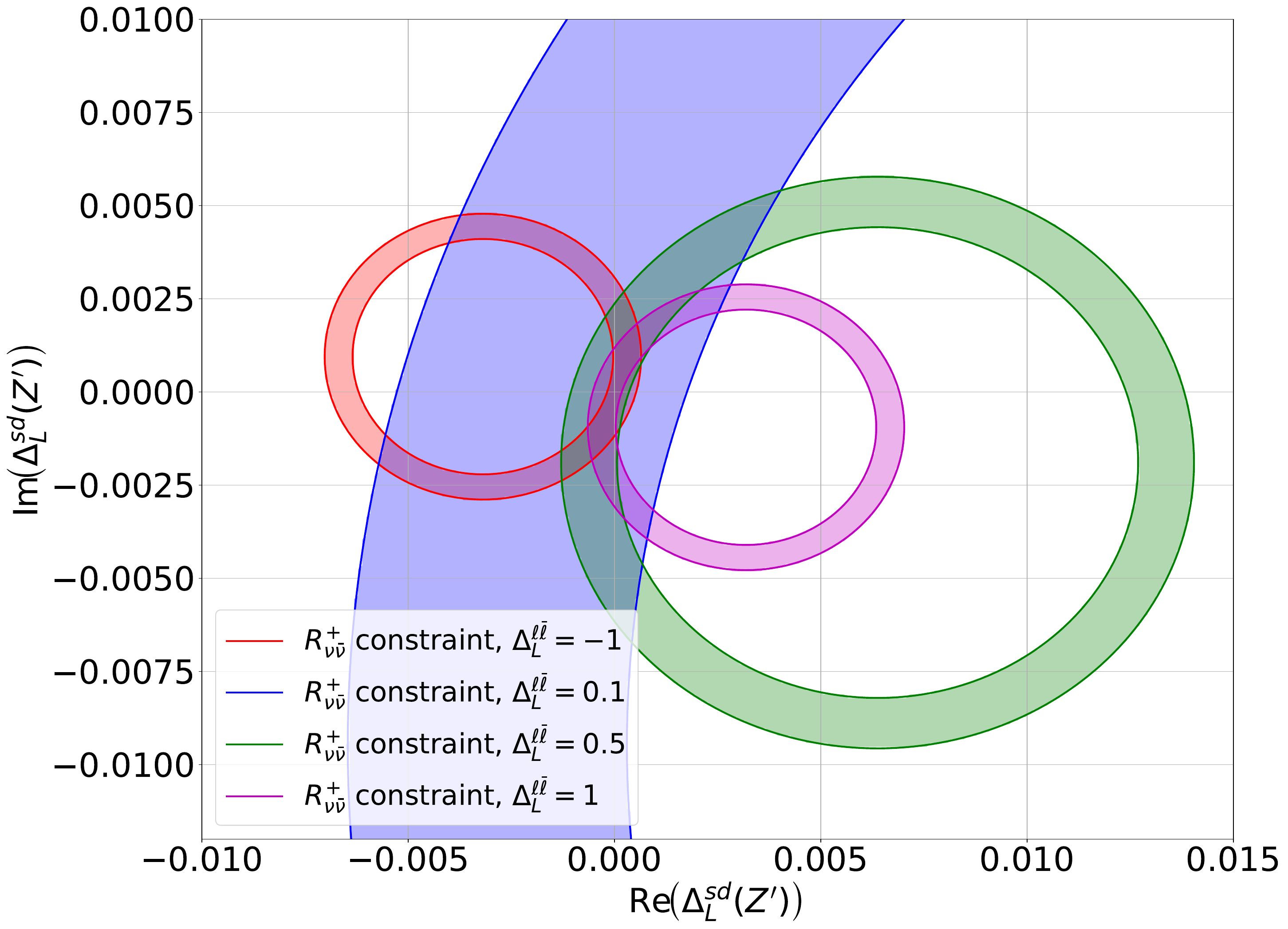}\hfill
    \includegraphics[width=0.48\linewidth]{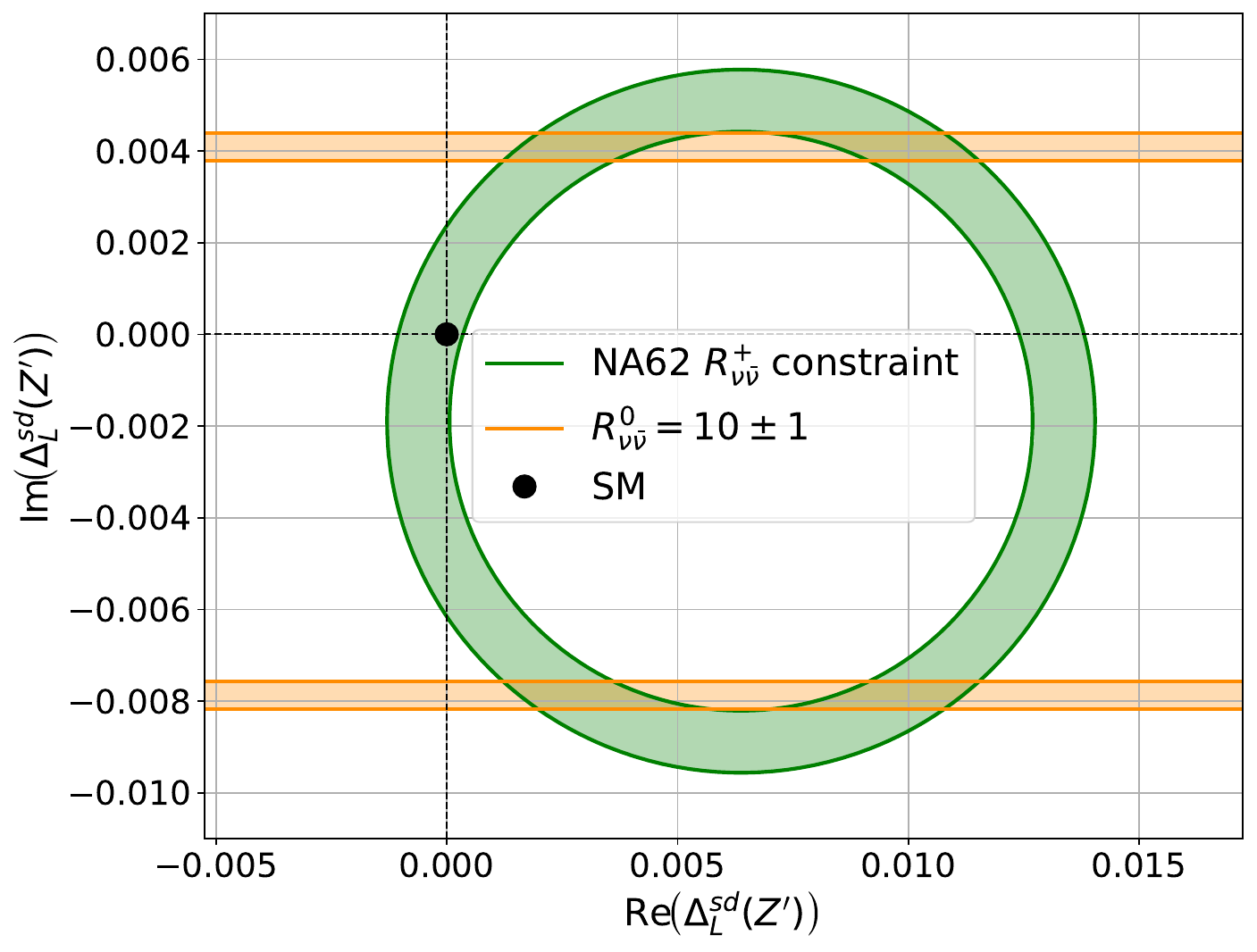}
    \caption{
    left: Constraints on $R_{\nu\bar{\nu}}^{+}$ and $R_{\nu\bar{\nu}}^{0}$ in the plane of $\left( \text{Re}\left(\Delta_{L}^{sd}(Z^{\prime})\right), \text{Im}\left(\Delta_{L}^{sd}(Z^{\prime})\right) \right)$ for different choices of $\Delta_{L}^{\ell\bar{\ell}}(Z')$.
    The shaded regions correspond to the allowed $1\sigma$ interval of the NA62 measurement.
    Right: Assuming $\Delta_{L}^{\ell\bar{\ell}}(Z')=0.5$, the NA62 constraint is shown in green and a hypothetical future measurement of $R^0_{\nu\bar{\nu}}=10\pm1$ is shown in orange. 
    When $\Delta_{L}^{sd}(Z^{\prime})=0$ there is no NP contribution and this therefore corresponds to the SM. 
    }
    \label{fig:R_nn_constraints}
\end{figure}

Regarding points 2 and 3, we have already seen in the analysis in Section~\ref{Interplay} that for particular values of the phase $\beta_X$ it is possible in the presence of a SM-like value for the $\kpn$ branching ratio to enhance the $\klpn$ branching ratio by an order of magnitude. 
Within this explicit $Z'$ scenario, the allowed phase-space depends on the product of 
$\Delta_{L}^{\ell\bar{\ell}}(Z^{\prime})\Delta_{L}^{sd}(Z^{\prime})$. 
The NA62-allowed region for $\Delta_L^{sd}(Z')$ is an annulus in the plane of 
$\left( \text{Re}\left(\Delta_{L}^{sd}(Z^{\prime})\right), \text{Im}\left(\Delta_{L}^{sd}(Z^{\prime})\right) \right)$, 
whose size depends on the chosen value of $\Delta_{L}^{\ell\bar{\ell}}$, as shown in the left panel of Fig.~\ref{fig:R_nn_constraints}.

The right panel of Fig.\ \ref{fig:R_nn_constraints} shows how a hypothetical future measurement of $R_{\nu\bar{\nu}}^{0} = 10\pm1$ by the KOTO II experiment would significantly constrain the allowed coupling parameters, with a residual four-fold degeneracy as discussed in Section~\ref{sec:KpnnInterplay}.
In what follows, we will assume $R_{\nu\bar{\nu}}^{+}=1$ and set $\Delta_L^{\ell\bar\ell}(Z') = 0.5$, leaving us with an allowed set of values for $\Delta_{L}^{sd}(Z^{\prime})$ which define a circle in the $\left( \text{Re}\left(\Delta_{L}^{sd}(Z^{\prime})\right), \text{Im}\left(\Delta_{L}^{sd}(Z^{\prime})\right) \right)$ plane.

Next on our list is the constraint on $\mathcal{B}(K_{L}\rightarrow\mu^{+}\mu^{-})_{\rm SD}$ from~\eqref{eq:KLmm-bound} which implies $R_{\mu^{+}\mu^{-}}^{L} < 3$, see point 4. 
$R_{\mu^{+}\mu^{-}}^{L}$ depends on the product $(\Delta_{R}^{\ell\bar{\ell}}(Z') - \Delta_{L}^{\ell\bar{\ell}}(Z')) \cdot \text{Re}\left(\Delta_{L}^{sd}(Z')\right)$.
Therefore, the requirement $R_{\mu^{+}\mu^{-}}^{L} < 3$ restricts the allowed parameter space in Fig.\ \ref{fig:R_nn_constraints} to a vertical band centred around $\text{Re}\left(\Delta_{L}^{sd}(Z')\right)=0$, whose width depends on the value of $(\Delta_{R}^{\ell\bar{\ell}}(Z') - \Delta_{L}^{\ell\bar{\ell}}(Z'))$.
In some coupling scenarios this constraint would therefore exclude two of the four possible solutions shown in the right panel of Fig.~\ref{fig:R_nn_constraints}.

All remaining observables on our list, 
namely $R_{\mu^+\mu^-}^S$ (point 5),
$R_{\pi\ell^{+}\ell^{-}}^{0}$ (point 6), 
and $\kappa_{\varepsilon^{\prime}}$ (point 7) 
are sensitive to CP-violating NP contributions and therefore depend on $\text{Im}\left(\Delta_{L}^{sd}(Z')\right)$. 
Yet they differ in their dependence on the flavour-conserving couplings $\Delta_{L,R}^{\ell\bar\ell}(Z')$ and $\Delta_R^{q\bar q}(Z')$.
The experimental constraints on ${\cal B}(\klpll)$ from the KTeV experiment (see~\eqref{KLLL}), combined with the SM predictions (see~\eqref{eq:KLpee} and~\eqref{eq:KLpmm}), correspond to constraints of
\begin{align}
    R_{\pi e^{+}e^{-}}^{0} < 8 \, (18)
    \qquad
    R_{\pi \mu^{+}\mu^{-}}^{0} < 27 \, (40)\,\,,
\label{eqn:RKleeConstraint}
\end{align}
if there is positive (negative) interference between the direct and mixing-induced contributions~\footnote{Both the numerator and denominator entering $R_{\pi \ell^{+}\ell^{-}}^{0}$ change between the positive and negative interference scenarios.}.

\begin{figure}
    \includegraphics[width=.49\textwidth]{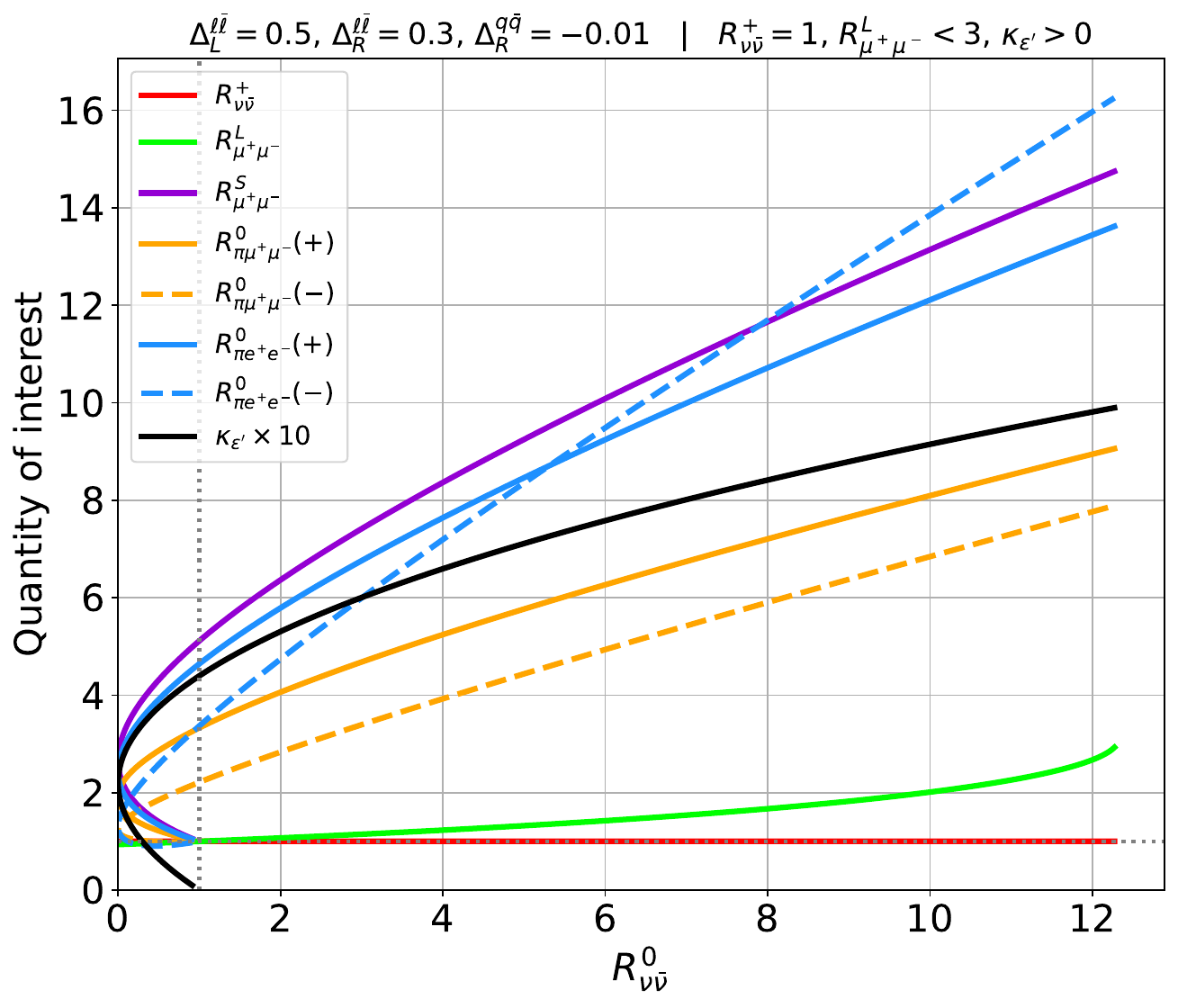}\hfill
    \includegraphics[width=.49\textwidth]{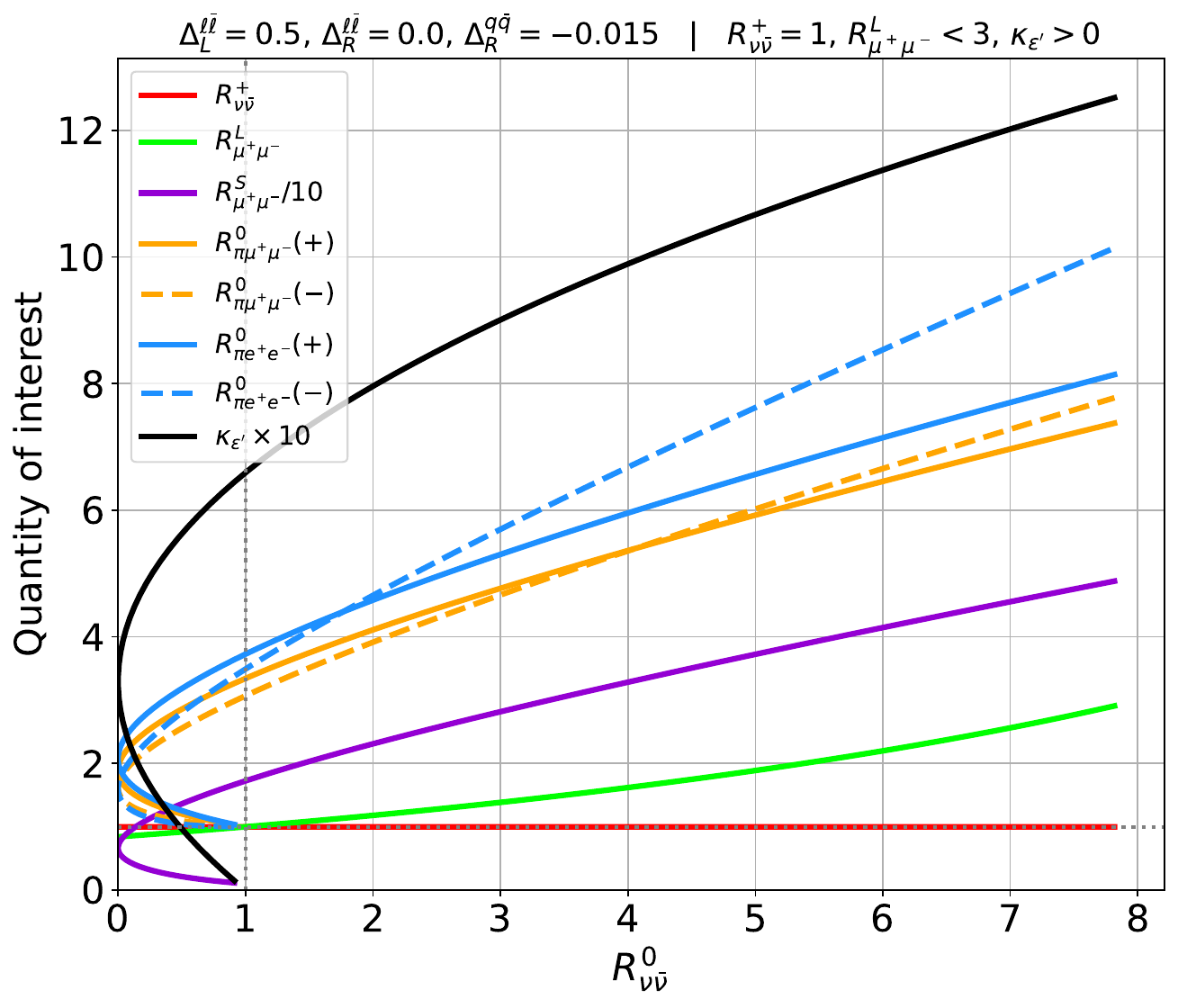}
    \caption{
$R_{\nu\bar\nu}^+$, $R_{\mu^+\mu^-}^{L,S}$, $R^{0}_{\pi\ell^{+}\ell^{-}}$ and $\kappa_{\varepsilon'}$
as functions of $R_{\nu\bar\nu}^0$ for two different choices of the flavour-conserving $Z'$ coupling parameters, 
enforcing $R_{\nu\bar\nu}^+ = 1$, $R_{\mu^+\mu^-}^L<3$ and $\kappa_{\varepsilon'} > 0$. 
For $R^{0}_{\pi\ell^{+}\ell^{-}}$, the solid (dashed) lines show the case of positive (negative) interference between direct and mixing-induced contributions.
Left:
    % Case 1
    $\Delta_{L}^{\ell\bar{\ell}}(Z^{\prime}) = 0.5$, 
    $\Delta_{R}^{\ell\bar{\ell}}(Z^{\prime}) = 0.3$, 
    $\Delta_R^{q\bar q}(Z') = -0.01$. 
Right:
    % Case 4
    $\Delta_{L}^{\ell\bar{\ell}}(Z^{\prime}) = 0.5$, $\Delta_{R}^{\ell\bar{\ell}}(Z^{\prime}) = 0$, 
    $\Delta_R^{q\bar q} (Z')= -0.015$.
}
    \label{fig:Ratios_master_plot}
\end{figure}

In Fig.\ \ref{fig:Ratios_master_plot} we show the quantities of interest (see~\eqref{eq:kappas}) 
$R_{\mu^+\mu^-}^{L,S}$, $R_{\pi\ell^{+}\ell^{-}}^{0}$ and $\kappa_{\varepsilon'}$
as functions of $R_{\nu\bar\nu}^0$ for two different choices of the flavour-conserving $Z'$ couplings ($\Delta_{L}^{\ell\bar{\ell}}(Z^{\prime})$, $\Delta_{R}^{\ell\bar{\ell}}(Z^{\prime})$ and $\Delta_{L}^{q\bar{q}}(R^{\prime})$),
as indicated in the caption. As discussed above, we enforce $R_{\nu\bar\nu}^+ = 1$---shown in red---which determines the shape of the curves. 
The extension of the curves to large $R_{\nu\bar\nu}^0$ values in the right panel is limited by the $R_{\mu^+\mu^-}^{L}$ bound. 
Finally, the requirement $\kappa_{\varepsilon'} \geq 0$ implies $\text{Im}\left(\Delta_{L}^{sd}(Z')\right) \geq 0$ for our choice of negative  $\Delta_R^{q\bar q}(Z')$. This determines the other endpoints of the curves.

We observe that in both cases all requirements (points 1--7) from our list are fulfilled simultaneously. In particular, in the scenario shown in the left panel of Fig.\ \ref{fig:Ratios_master_plot}, the Grossman-Nir bound for $\klpn$ can be reached and simultaneously all other observables can be significantly enhanced. 
When assuming positive interference between the direct and mixing-induced contributions to $K_L\to \pi^0e^+e^-$ we find $R_{\pi e^{+}e^{-}}^{0}(+)>8$ for $R_{\nu\bar\nu}^0\gsim 4.5$, which is excluded experimentally according to \eqref{eqn:RKleeConstraint}. In the negative interference scenario the ratio $R_{\pi e^{+}e^{-}}^{0}(-)$ remains below the experimental constraint for the full range of allowed $R_{\nu\bar\nu}^{0}$.
(See also the $\Delta_{R}^{\ell\bar{\ell}}(Z^{\prime})=0.3$ curves in the top panel of Figure~\ref{fig:BRKLpiee_vs_RSmm} below.)
Yet, even in the negative-interference case much larger decay rates than predicted in the SM are expected.

Specifically, for the coupling parameters shown in the left panel of Fig.\ \ref{fig:Ratios_master_plot}, when $R_{\nu\bar\nu}^0$ reaches the Grossman-Nir bound while $R_{\nu\bar\nu}^+=1$, the bound $R_{\mu^+\mu^-}^L <3$ is saturated. 
At the same time, the branching ratios ${\cal B}(K_L\to\pi^0e^+e^-) = 25\times 10^{-11}$ and ${\cal B}(K_L\to\pi^0\mu^+\mu^-) = 7.5 \times 10^{-11}$ are predicted for the required negative interference. 
The ratio $R_{\mu^+\mu^-}^S$ reaches values as high as 15 while $\kappa_{\varepsilon'}$ can be $\ord(1)$. 
Note that in the future, once the required amount of NP in $\epe$ is more precisely known, the size of the $Z'$ contribution can be changed by adjusting the value of $\Delta_R^{q\bar q}(Z')$ accordingly, without affecting any of the other results presented here.

Comparing the two coupling scenarios shown in Fig.\ \ref{fig:Ratios_master_plot}, we identify some differences in the quantitative pattern of effects. 
The vanishing $Z'$ coupling to RH leptons in the right panel leads to a significantly larger axial-vector coupling $\Delta_A^{\ell\bar\ell}(Z')$ entering the $K_{L,S}\to \mu^+\mu^-$ decays. 
The bound on $R_{\mu^+\mu^-}^L$ therefore does not leave enough room for $\text{Re}\left(\Delta_{L}^{sd}(Z')\right)$ to let $\klpn$ reach the Grossman-Nir bound. 
However, large enhancements of that branching ratio, up to almost a factor of 8, are still possible. 
At the same time, both positive and negative interference patterns remain allowed for $\klpll$ and both branching ratios can be significantly enhanced. 
We also notice that $R_{\mu^+\mu^-}^S$ can grow much larger in the right panel which is again due to the larger axial-vector coupling $\Delta_A^{\ell\bar\ell}(Z')$. 
Finally the larger value of $|\Delta_R^{q\bar q}(Z')|$ implies again possible $\ord(1)$ values of $\kappa_{\epsilon'}$ despite the smaller $\text{Im}\left(\Delta_L^{sd}(Z')\right)$, as expected from the linear dependence of $\kappa_{\varepsilon^{\prime}}$ on the $\Delta_R^{q\bar q}(Z')$ coupling parameter.

\begin{figure}
\hfill
\centering{\includegraphics[width=.49\textwidth]{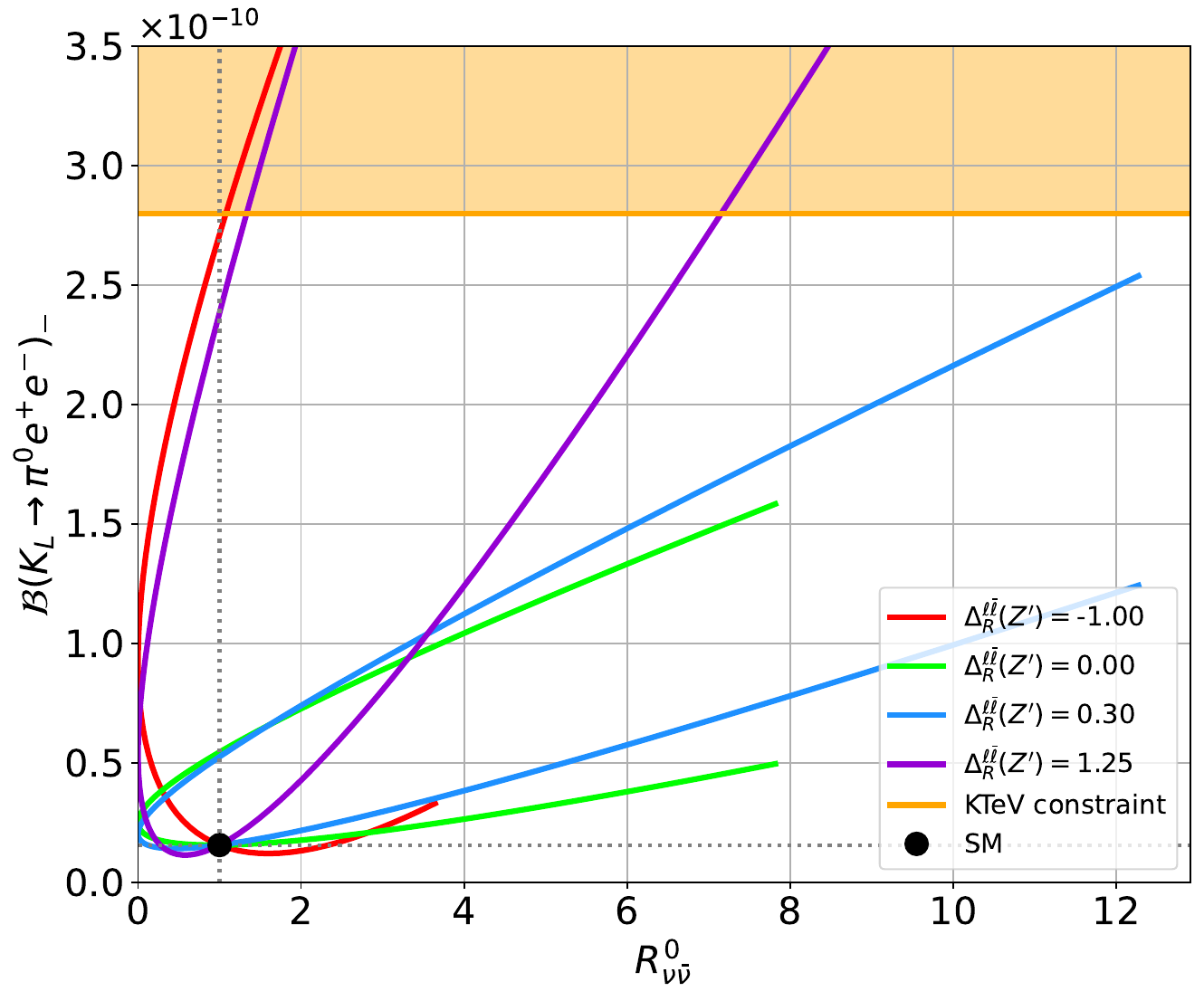}}
\hfill
\centering{\includegraphics[width=.49\textwidth]{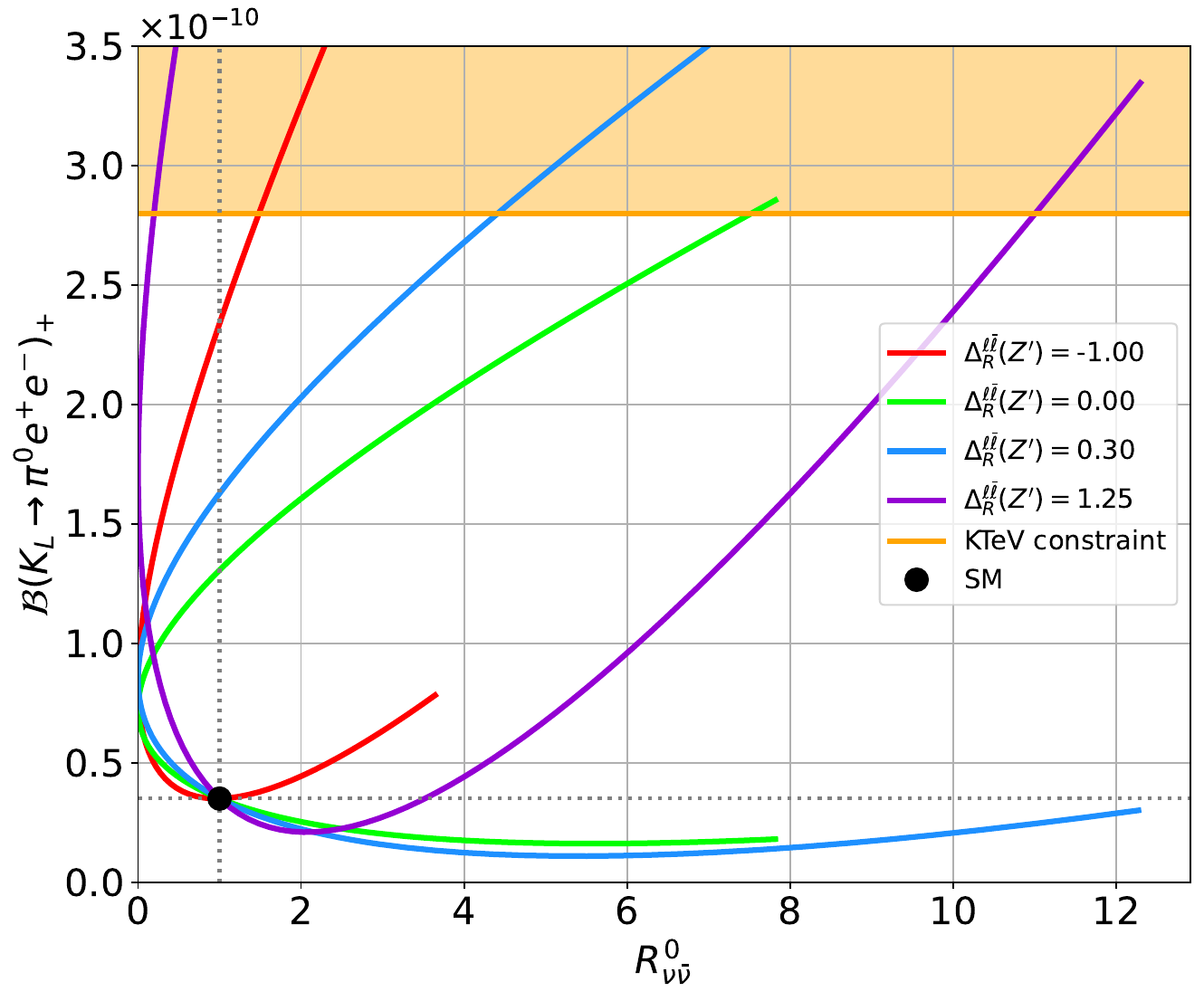}}
\hfill
\centering{\includegraphics[width=.49\textwidth]{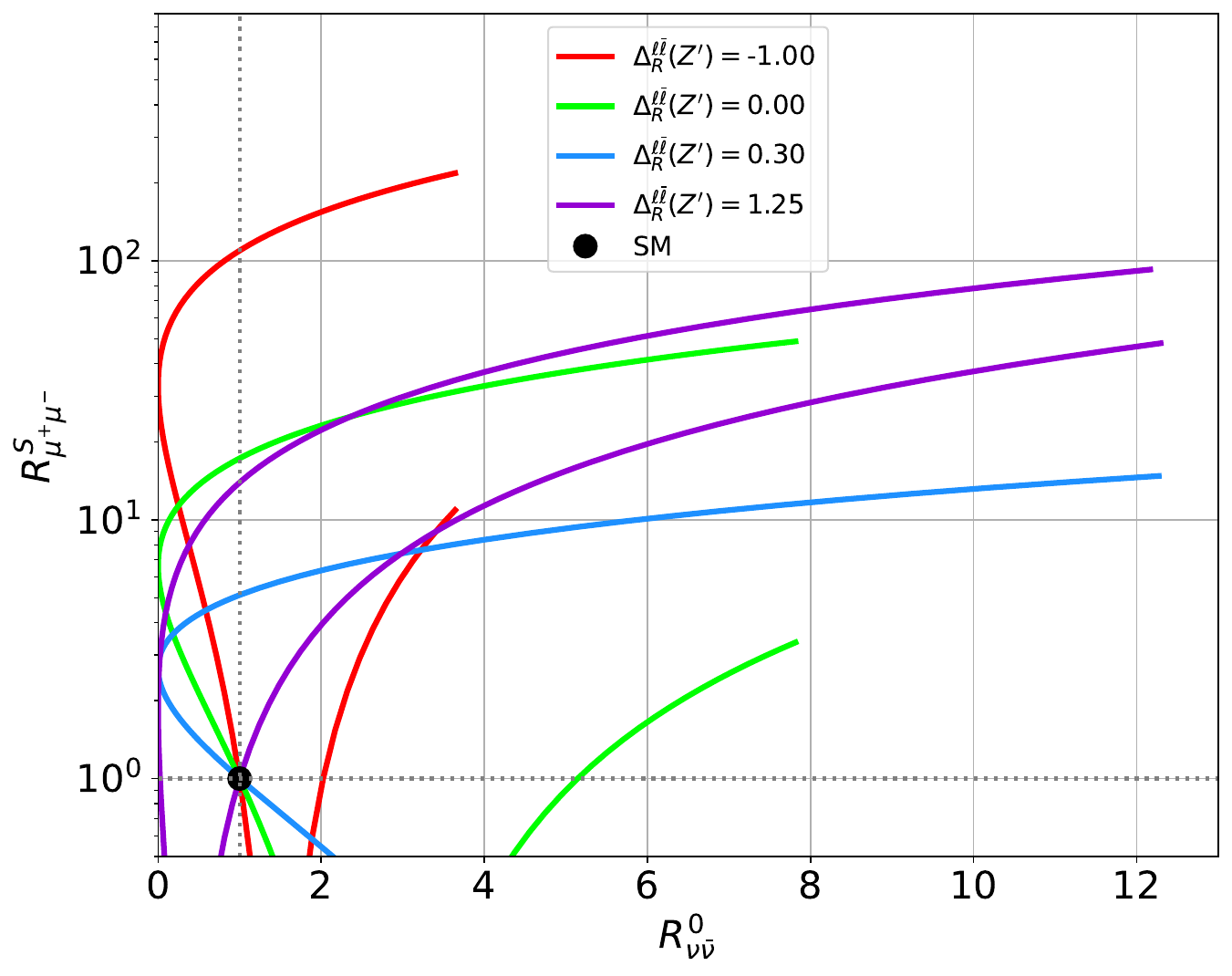}}
\hfill
\centering{\includegraphics[width=.49\textwidth]{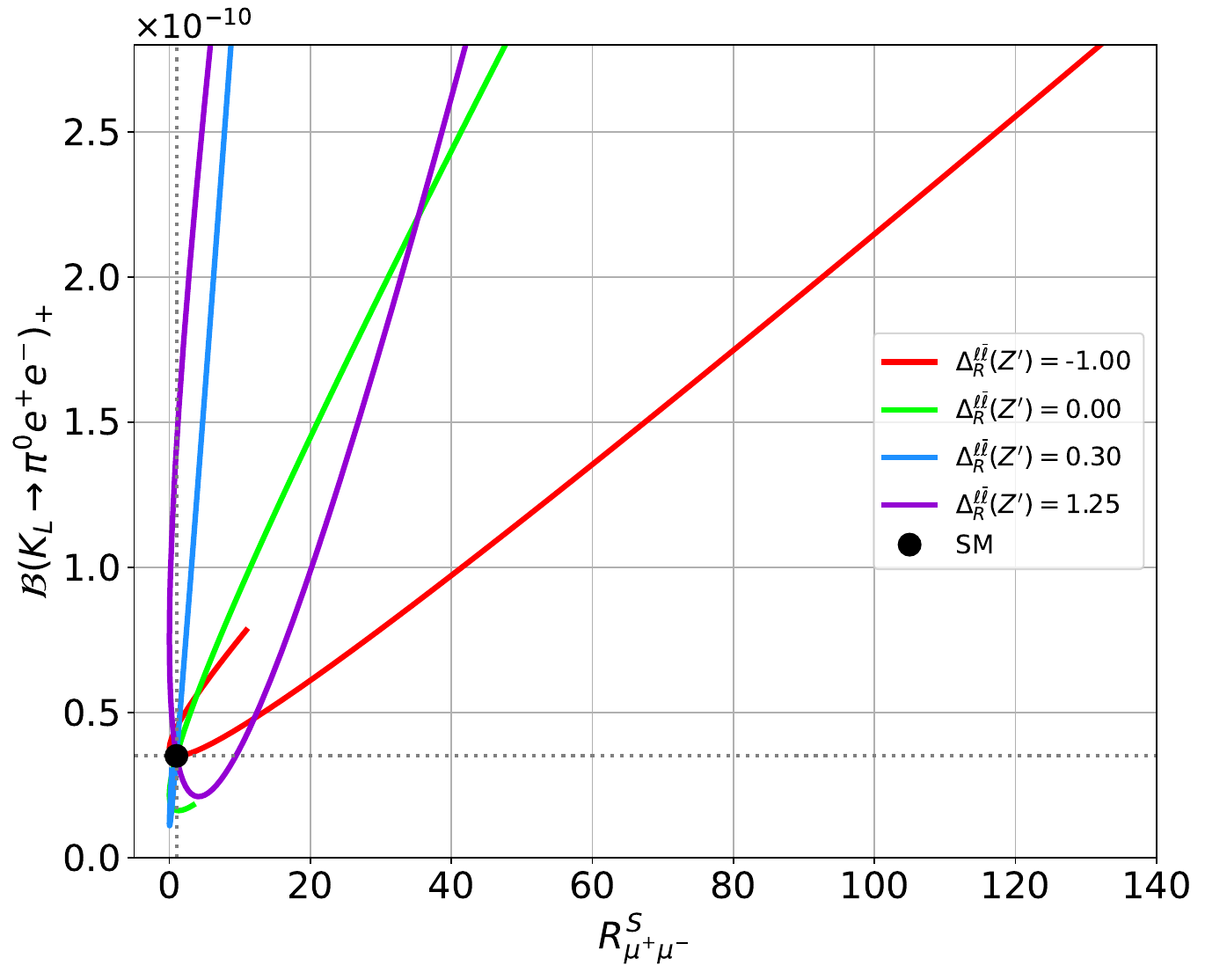}}
\caption{
Top: comparison of $\mathcal{B}(K_{L}\rightarrow\pi^{0}e^{+}e^{-})$ versus $R_{\nu\bar{\nu}}^{0}$ when assuming negative (left) and positive (right) interference. The experimental constraint on $\mathcal{B}(K_{L}\rightarrow\pi^{0}e^{+}e^{-})<2.8\times10^{-11}$ from the KTeV experiment~\cite{AlaviHarati:2003mr} (see eq.~\eqref{KLLL}) is shown in orange, where the shaded region is excluded at $90\%$ C.L..
Lower left: study of $R_{\mu^{+}\mu^{-}}^{S}$ versus $R_{\nu\bar{\nu}}^{0}$.
Lower right: 
Correlation between the $R_{\mu^{+}\mu^{-}}^{S}$ variable and the branching ratio $\mathcal{B}(K_{L}\rightarrow\pi^{0}e^{+}e^{-})$ (assuming positive interference). %for $\Delta L^{\ell\ell}(Z')=0.5$.
In all cases $\Delta_{L}^{\ell\bar{\ell}}(Z^{\prime})=0.5$ is used, 
the constraints $R_{\nu\bar{\nu}}^{+}=1$ and $R_{\mu^{+}\mu^{-}}^{L}<3$ are imposed, 
and results using several values of $\Delta_{R}^{\ell\bar{\ell}}(Z^{\prime})$ are shown.
}
\label{fig:BRKLpiee_vs_RSmm}
\end{figure}

A more detailed investigation of the possible enhancements of the rates of the $K_{S}\rightarrow\mu^{+}\mu^{-}$ and $K_{L}\rightarrow\pi^{0}e^{+}e^{-}$ decays, of interest for future study at the LHCb and KOTO II experiments, respectively, are shown in Fig.~\ref{fig:BRKLpiee_vs_RSmm}. 
Keeping $\Delta_{L}^{\ell\bar{\ell}}(Z^{\prime})=0.5$, and imposing the constraints $R_{\nu\bar{\nu}}^{+}=1$ and $R_{\mu^{+}\mu^{-}}^{L}<3$,
various values of $\Delta_{R}^{\ell\bar{\ell}}(Z^{\prime})$ are considered. 
In the top panel a comparison of $\mathcal{B}(K_{L}\rightarrow\pi^{0}e^{+}e^{-})$ assuming positive or negative interference between direct and mixing-induced contributions (see~\eqref{eq:BrKpiLL}) is shown: this highlights the significant differences in the branching ratio both for different signs of the interference and different choices of $\Delta_{R}^{\ell\bar{\ell}}(Z^{\prime})$. 
In the lower left panel the values of $R_{\mu^{+}\mu^{-}}^{S}$ versus $R^{0}_{\nu\bar{\nu}}$ are shown. We find possible values of $R_{\mu^{+}\mu^{-}}^{S}$ of more than $100$, well within the range of the future sensitivity of LHCb.
The lower-right panel then shows the correlation between $\mathcal{B}(K_{L}\rightarrow\pi^{0}e^{+}e^{-})$ (assuming positive interference) and $R_{\mu^{+}\mu^{-}}^{S}$.
With all considered choices of $\Delta_{R}^{\ell\bar{\ell}}(Z^{\prime})$
a significant enhancement of $\mathcal{B}(K_{L}\rightarrow\pi^{0}e^{+}e^{-})$ is possible, saturating the experimental constraint which corresponds to the maximum vertical extent of the lower-right panel. In general $K_{S}\rightarrow\mu^{+}\mu^{-}$ is also enhanced substantially. 
We find that a scenario where only $K_{S}\rightarrow\mu^{+}\mu^{-}$ is enhanced and $K_{L}\rightarrow\pi^{0}e^{+}e^{-}$ is SM-like is 
disfavoured in this particular model, given the existing experimental constraints. 
However, the opposite case, $K_{S}\rightarrow\mu^{+}\mu^{-}$ being SM-like while $K_{L}\rightarrow\pi^{0}e^{+}e^{-}$ is significantly enhanced, is possible.
Therefore, studies of both of these decays are essential.

\section{Summary}\label{sec:5}

Rare and CP-violating kaon decays are among the most promising targets in the search for NP at high scales. 
We have demonstrated in the present article that this statement remains true after the recent NA62 measurement of ${\cal B}(\kpn)$ consistent with the SM expectation, and despite the good agreement of the $K^0-\bar K^0$ mixing observables with their SM predictions.
While at first sight the agreement of the NA62 result for ${\cal B}(\kpn)$ with its SM prediction may seem disappointing, it is a very important result with profound implications for the pattern of possible NP effects in other rare kaon decays.

We have first investigated the impact of NP contributions on the $K\to\pi\nu\bar\nu$ decays in a model-independent analysis, stressing that in order to achieve values of ${\cal B}(\klpn)$ near the Grossman-Nir bound, a large phase $\beta_X \approx 110^\circ$ of the relevant function $X_\text{eff}$ is required. 
We have also demonstrated how measurements of the $\kpn$ and $\klpn$ branching ratios by NA62 and KOTO II, respectively, can be used to determine the values of  $|X_\text{eff}|$
and $\beta_X$.

We have then turned our attention to a specific $Z'$ scenario with flavour-violating $sd$ couplings, in order to correlate the possible NP pattern in $K\to\pi\nu\bar\nu$ with other kaon decays of interest, in particular the neutral kaon mixing observables $\varepsilon_K$ and $\Delta M_K$, the branching ratios of the decays $K_{L,S}\to\mu^+\mu^-$ and $\klpll$, as well as the parameter $\epe$ measuring direct CP violation in $K\to\pi\pi$.
Our analysis proves that large enhancements of the decays $\klpn$, $\klpll$ and $\ksm$ as well as of the parameter $\epe$ remain simultaneously possible despite the $K^0-\bar K^0$ observables being in good agreement with the SM prediction and the constraint from $K_L\to\mu^+\mu^-$, even in the most pessimistic scenario of no NP in the decay $\kpn$.

It should be emphasised that the $Z^\prime$ scenario considered by us is only
an illustration of a particular NP scenario at work, and other more complete NP models may predict somewhat different patterns of effects.

Our work thus strongly encourages the continuation of the experimental kaon physics programme, with future measurements of the decays $\klpn$, $K_L \to \pi^0e^+e^-$ and possibly $K_L\to\pi^0\mu^+\mu^-$ by the KOTO II experiment as well as of $\ksm$ by LHCb, to be complemented by continued efforts by the theory and lattice communities to improve the SM prediction of $\epe$.
On the theory side, the NA62 result already has an impact on the correlations
between rare kaon and $B$ meson decays in concrete NP scenarios, as
evident from the DNA-charts in \cite{Buras:2013ooa}. Indeed, the first study
of the implications of the NA62 result on such correlations appeared
recently \cite{Allwicher:2026loe}. Yet, the full assessment will only
be possible when the data on rare $B$ and $K$ decays improves in the coming years.

%%%%%%%%%%%%%%%%%%%%%%%%%%%%%%%%%%%%%%%%%%%%%%%%%%%%%%%%%%%%%%%%%%%%%%
\section*{Acknowledgements}
The work of M.B. is supported by the DFG under grant 396021762 -- TRR~257. M.B. is grateful to the Mainz Institute for Theoretical Physics (MITP) of the Cluster of Excellence PRISMA+ (Project ID 390831469) for its hospitality and partial support during parts of this work.
A.J.B. acknowledges financial support from the Excellence Cluster ORIGINS,
funded by the Deutsche Forschungsgemeinschaft (DFG, German Research Foundation)
under Germany's Excellence Strategy – EXC-2094 – 390783311.
The work of C.L. is partially funded by STFC under grant UKRI/ST/C002848/1.

%%%%%%%%%%%%%%%%%%%%%%%%%%%%%%%%%%%%%%%%%%%%%%%%%%%%%%%%%%%%%%%%%%%%%%
\vspace{+0.5cm}
\appendix 
%\vspace{+0.25cm}

\section{CKM Parameters from BV Strategy}\label{CKM}
Basing the determination of CKM parameters exclusively on $\Delta F=2$ observables as done in \cite{Buras:2022wpw} and 
updating the value of $\beta=22.62(45)^\circ$ from HFLAV \cite{HeavyFlavorAveragingGroupHFLAV:2024ctg} relative to $\beta=22.2(7)$ used in \cite{Buras:2022wpw}
one finds \cite{Buras:2026vbp}
\be\label{CKMBV}
{\vcb=42.5(5)\times 10^{-3}, \quad 
  \gamma=64.6(16)^\circ, \quad \beta=22.62(45)^\circ, \quad \vub=3.76(11)\times 10^{-3}\,}
\ee
and consequently 
\be\label{CKMoutput2}
{\vts=41.8(4)\times 10^{-3}\,, \qquad \beta_{s} = -1.05(4)^{\circ}\,, \qquad {\vtd=8.65(15)\times 10^{-3}}\,.}
\ee
{We then have}
\be
{{\RE}\lambda_c = -0.2189(9)\,,\quad
{\RE}\lambda_t = -3.31(10) \times 10^{-4}\,,\quad}
{{\IM}\lambda_t=1.44(5)\times 10^{-4}\,,}
\ee
{where $\lambda_c=V_{cs}^*V_{cd}$ and $\lambda_t=V_{ts}^*V_{td}$.}
%, as well as
%\be\label{CKMoutput3}
%{\bar\varrho=0.164(11)\,,\qquad \bar\eta=0.346(12)\,.}
%\ee}
%where $\lambda_t=V_{ts}^*V_{td}$.

%%%%%%%%%%%%%%%%%%%%%%%%%%%%%%%%%%%%%%%%%%%%%%%%%%%%%%%%%%%%%%%%%%%%%%
\addcontentsline{toc}{section}{References}

\small

\bibliographystyle{JHEP}
\bibliography{Bookallrefs}

\end{document}